\documentclass[aps,twocolumn,prd,superscriptaddress,longbibliography,reprint]{revtex4}
\usepackage{amsfonts}
\usepackage{mathrsfs}
\usepackage{amsmath}
\usepackage{color}
\usepackage{graphicx}
\usepackage{bm}
\usepackage{amssymb}
\usepackage{xspace}
\usepackage{epstopdf}
\usepackage{dcolumn}
\usepackage{longtable}
\usepackage{multirow}
\usepackage[colorlinks=true, letterpaper=true, pdfstartview=FitV, linkcolor=blue, citecolor=blue, urlcolor=blue]{hyperref}

\begin{document}

\title{Graphyne as a second-order and real Chern topological insulator in two dimensions}

\author{Cong Chen}
\affiliation{School of Physics, and Key Laboratory of Micro-nano Measurement-Manipulation and Physics (Ministry of Education), Beihang University, Beijing 100191, China}
\address{Research Laboratory for Quantum Materials, Singapore University of Technology and Design, Singapore 487372, Singapore}

\author{Weikang Wu}
\address{Research Laboratory for Quantum Materials, Singapore University of Technology and Design, Singapore 487372, Singapore}

\author{Zhi-Ming Yu}
\affiliation{Key Lab of Advanced Optoelectronic Quantum Architecture and Measurement (MOE),
Beijing Key Lab of Nanophotonics and Ultrafine Optoelectronic Systems, and School of Physics,
Beijing Institute of Technology, Beijing 100081, China}
\address{Research Laboratory for Quantum Materials, Singapore University of Technology and Design, Singapore 487372, Singapore}

\author{Ziyu Chen}
\affiliation{School of Physics, and Key Laboratory of Micro-nano Measurement-Manipulation and Physics (Ministry of Education), Beihang University, Beijing 100191, China}

\author{Y. X. Zhao}
\email{zhaoyx@nju.edu.cn}
\address{National Laboratory of Solid State Microstructures and Department of Physics, Nanjing University, Nanjing 210093, China}
\address{Collaborative Innovation Center of Advanced Microstructures, Nanjing University, Nanjing 210093, China}

\author{Xian-Lei Sheng}
\email{xlsheng@buaa.edu.cn}
\affiliation{School of Physics, and Key Laboratory of Micro-nano Measurement-Manipulation and Physics (Ministry of Education), Beihang University, Beijing 100191, China}
\address{Research Laboratory for Quantum Materials, Singapore University of Technology and Design, Singapore 487372, Singapore}

\author{Shengyuan A. Yang}
\address{Research Laboratory for Quantum Materials, Singapore University of Technology and Design, Singapore 487372, Singapore}

\begin{abstract}
Higher-order topological phases and real topological phases are two emerging topics in topological states of matter, which have been attracting considerable research interest. However, it remains a challenge to find realistic materials that can realize these exotic phases. Here, based on first-principles calculations and theoretical analysis, we identify graphyne, the representative of the graphyne-family carbon allotropes, as a two-dimensional (2D) second-order topological insulator and a real Chern insulator. We show that  graphyne has a direct bulk band gap at the three $M$ points, forming three valleys. The bulk bands feature a double band inversion, which is characterized by the nontrivial real Chern number enabled by the spacetime-inversion symmetry. The real Chern number is explicitly evaluated by both the Wilson-loop method and the parity approach, and we show that it dictates the existence of Dirac type edge bands and the topological corner states. Furthermore, we find that the topological phase transition in graphyne from the second-order topological insulator to a trivial insulator is mediated by a 2D Weyl semimetal phase. The robustness of the corner states against symmetry breaking and possible experimental detection methods are discussed.
\end{abstract}

\maketitle

\section{Introduction}

Graphyne and graphyne-family members are two-dimensional (2D) carbon allotropes consisting of $sp$ and $sp^2$-bonded carbon atoms, originally proposed by Baughman \emph{et al.} in 1987~\cite{GDY1987}. Like graphene, these materials are completely flat with single-atom-thickness, and their structures can be viewed as derived from graphene by replacing some of the carbon bonds with acetylenic groups~\cite{Diederich1994uk,ZhangJin_ChemSocRev,LiYL2010}. In the past several decades, especially after the realization of graphene, these 2D carbon materials have been attracting tremendous research interest. In 2010, a member of the graphyne family, graphdiyne, was successfully synthesized by Li \emph{et al}~\cite{LiYL2010}. Meanwhile, graphyne (sometimes also known as graphyne-1 or $\gamma$-graphyne~\cite{GDY1987,Diederich1994uk,ZhangJin_ChemSocRev,Bunz1999}; here we follow Baughman \emph{et al.}'s original notation to simply refer to it as graphyne)~\cite{GDY1987}, as the representative member of the family, has also been demonstrated in the form of small fragments in experiment~\cite{Diederich1994uk,Bunz1999}.

Previous studies on graphyne and graphyne-family materials have been mostly focused on their excellent mechanical~\cite{Ajori2013,Babak2020}, electronic~\cite{Li2015nc,ShuaiZG2013jpcl}, catalytic~\cite{Yang2013cm,WangD2014,Xue2018vq}, and thermal properties~\cite{Zhao2015ComMatSci,Tao_PhysRevB2012,Solis_ACS2019}, while less attention is paid on their topological property. The main reason is that the spin-orbit coupling (SOC), which is required for \emph{conventional} topological insulators (TIs), is negligibly small in carbon materials. Similar to graphene, although some of the graphyne-family members host Fermi points in their band structure~\cite{Malko_PRL,NatSciRev2015}, SOC is too weak to open a detectable gap at these points to realize a 2D TI (also known as the quantum spin Hall insulator)~\cite{Hongki_PRB,YuguiYao_PRB}. This is especially true for graphyne, because it is a semiconductor with a sizable band gap $\sim 0.5$ eV~\cite{Narita1998}, ruling out any possibility for an SOC-dominated band gap necessary for a 2D TI.

Nevertheless, the above argument does not exclude the possibility of a higher-order band topology, which can be achieved in the absence of SOC.
As an extension of the TI concept, an $n$-th order TI in $d$ spatial dimensions features protected gapless states at its $(d-n)$-dimensional boundary, but is gapped otherwise~\cite{ZhangFan_PRL2013,Hughes2017,Langbehn2017,SongZD2017,Schindler2018SA}. Hence, a second-order TI (SOTI) in 2D will have topologically protected gapless states at its 0D corners, while its 2D bulk and 1D edges are generically insulating. So far, the SOTI phase has been mostly explored in artificial periodic systems~\cite{Kruthoff2017,Ezawa2018L,Imhof2018wj,SerraGarcia2018,Peterson2018,Yuhan_arXiv,Xue2019to,JiangH2020}. As for real materials, in 3D, a few candidates such as SnTe~\cite{Schindler2018SA}, bismuth~\cite{Schindler2018}, $X$Te$_2$ ($X=$ Mo, W)~\cite{ZJWang_PRL2019}, Bi$_{2-x}$Sm$_x$Se$_3$~\cite{Yue2019ws}, EuIn$_2$As$_2$~\cite{XuYF2019}, and MnBi$_{2n}$Te$_{3n+1}$~\cite{MnBiTe_PRL}, have been proposed. In 2D, the identified real materials are even less, which poses a challenge for subsequent studies.

In a previous work, we identified graphdiyne as the first realistic 2D SOTI~\cite{GDY_PRL2019}, which features a double band inversion at the $\Gamma$ point, gapped Dirac-type edge spectrum, and robust topological corner states. The essential result is also confirmed in Ref.~\cite{Lee_GDY}. Since graphyne is the paradigmatic example of the graphyne-family with an even simpler structure than graphdiyne, one may naturally wonder: \emph{Is there also a nontrivial topology in graphyne?}

In this work, we explore the answer to this question. We find that graphyne is also a 2D SOTI, and more specifically, a 2D real Chern insulator (RCI) characterized by a nontrivial real Chern number~\cite{RealCN_YXZhao}. Via combined first-principles calculations and theoretical analysis, we show that graphyne has a triple-valley structure with direct band gap at the three $M$ points of the Brillouin zone (BZ), and there is a double band inversion feature. While both its bulk and 1D edges are gapped, a pair of nontrivial edge bands appear on each edge. We show that these edge bands are dictated by the nontrivial bulk real Chern number $\nu_R$ and must take a 1D Dirac form. The real Chern number is enabled by the spacetime inversion symmetry $\mathcal{PT}$, and is explicitly evaluated here by using both the Wilson-loop method and the parity approach. It follows that each edge is an 1D Dirac insulator with a $\mathbb{Z}_2$ topological classification. At a corner where two edges with different $\mathbb{Z}_2$ characters meet, there must exist a 0D topological corner state, corresponding to the topological domain-wall mode. The topological phase transitions under strain and the robustness of the SOTI against symmetry breaking are discussed.

Compared with our previous work on graphdiyne~\cite{GDY_PRL2019}, we wish to emphasize the following distinctions of the current work. First, their band structures are very different. For graphyne, the band gap (and the low-energy physics) is at the three $M$ points, not at the $\Gamma$ point as in graphdiyne. Consequently, their edge spectra also exhibit different patterns. Second, for graphyne, we reveal an interesting topological phase transition not observed before, where the transition between a SOTI and a trivial insulator is mediated by a 2D Weyl semimetal phase. Third, the bulk topological invariant, namely, the real Chern number, and its connection to the boundary modes are discussed in much more detail in the current work. Finally, in this work, we explicitly demonstrate that the topological corner states are robust against perturbations that break the symmetries (such as the inversion). This will greatly broaden the experimental relevance of the topological phase in graphyne.

\begin{figure}
  \includegraphics[width=8 cm]{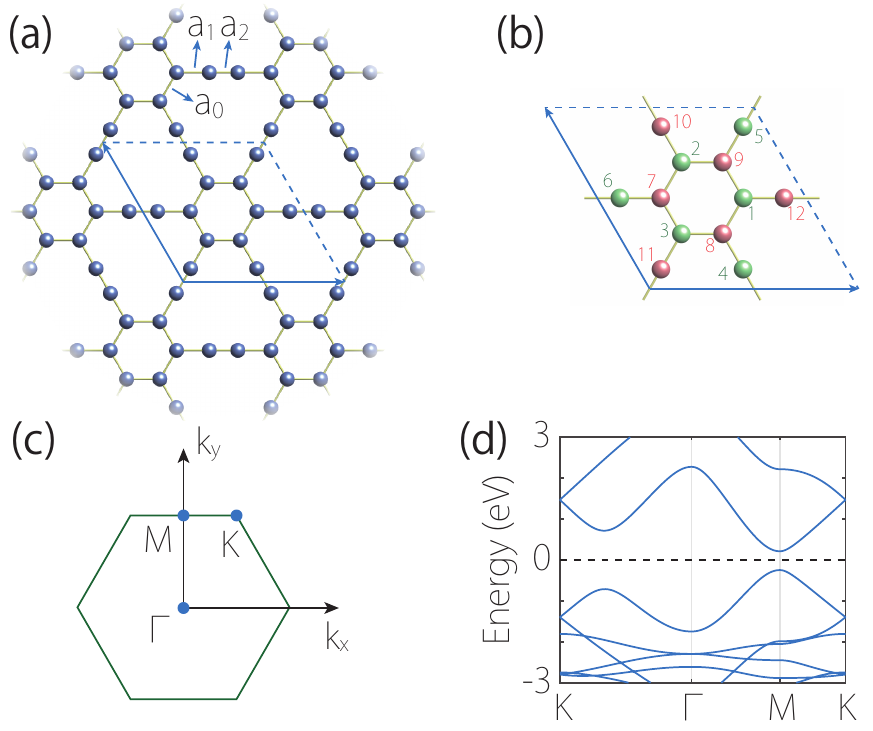}
  \caption{(a) Crystal structure of graphyne. (b) Unit cell of graphene. It has a bipartite lattice, namely, the atomic sites can be divided into two sublattices, as marked by the two colors in the figure. When hopping between distinct sublattices dominates, the system preserves the sublattice symmetry. (c) Brillouin zone with high symmetry points labeled. (d) Bulk electronic band structure of graphyne. The direct band gap occurs at the three $M$ points.}
\label{crystal}
\end{figure}

\section{Crystal structure}

Graphyne has a completely planar lattice structure with a single-atom thickness. As shown in Fig.~\ref{crystal}(a), the structure can be constructed by inserting one acetylenic linkage between two nearby hexagonal carbon rings in the graphene lattice. It has been shown that graphyne enjoys a high thermal stability, and it is the most stable member in the whole graphyne-family~\cite{Solis_ACS2019}.
The optimized lattice constant from our first-principles calculations is 6.890~\AA\ (the calculation details are presented in  Supplemental Material), which is in agreement with the previous study~\cite{Narita1998,JunKangJPCC}. It has the same hexagonal crystalline symmetry $p$6m as graphene and graphdiyne. Each unit cell contains 12 atoms, which can be divided into two sublattices, as indicated in Fig.~\ref{crystal}(b). This gives rise to an emergent sublattice (chiral) symmetry $\mathcal{C}$ for the low-energy bands, as we shall discuss later. The other two symmetries which are important in our analysis are the inversion $\mathcal{P}$ and time reversal $\mathcal{T}$ symmetries.
There are three nonequivalent bonds in the graphyne lattice, as marked in Fig.~\ref{crystal}(a).
The calculated values are $a_0$= 1.426~\AA, $a_1$= 1.408~\AA, and $a_2$= 1.222~\AA, also in agreement with previous studies~\cite{Narita1998,JunKangJPCC}.

\section{Bulk band structure}

Figure~\ref{crystal}(d) shows the calculated band structure of graphyne. One observes that graphyne is a direct band gap semiconductor with the gap located at the $M$ point (there are three $M$ points in the BZ related by the $C_6$ symmetry). The band gap is about 0.46 eV on the GGA level. Note that the SOC effect is negligible here since carbon is a light element.
We have checked the band structure with SOC, which shows little difference. Therefore, in the DFT calculations as well as the following analysis, SOC and hence the spin degree of freedom will be disregarded, and the system is effectively spinless.

Considering the possible underestimation of band gap by GGA, we have further checked the band structure by using the more accurate modified Becke-Johnson potential (see  Supplemental Material). The result shows that the bandgap is slightly increased to $\sim$0.53 eV, while all essential band features remain the same.


\begin{figure}[t!]
\includegraphics[width=8 cm]{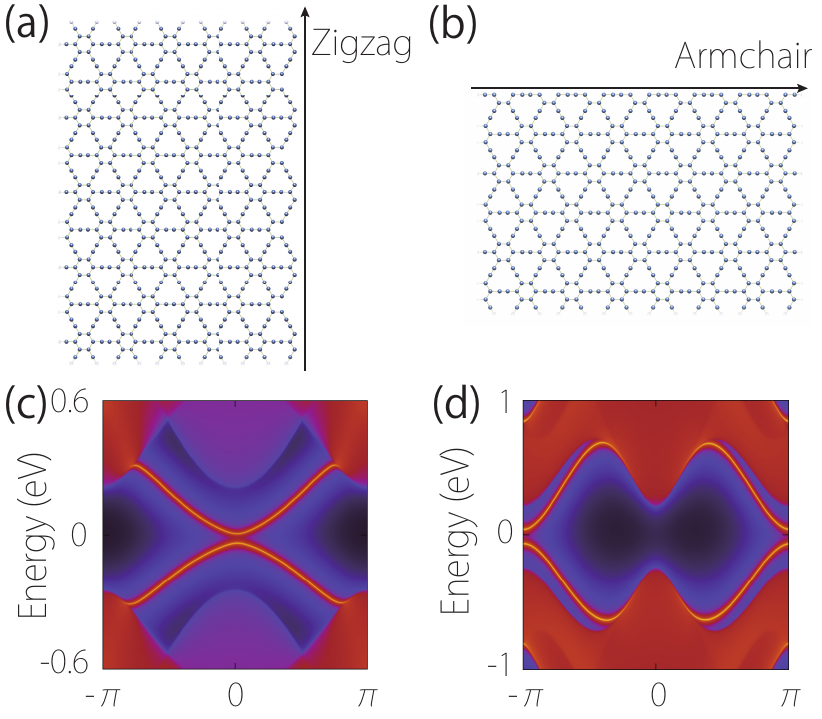}
\caption{A semi-infinite graphyne with (a) zigzag and (b) armchair edge. (c) and (d) show the projected spectra for the zigzag edge and the armchair edge, respectively.}
\label{edge}
\end{figure}

As we have discussed, graphyne cannot be a conventional 2D TI. However, a clue of nontrivial band topology can be observed from the symmetry properties of the occupied (valence) bands. Consider the $\mathcal{P}$ parity eigenvalues for the valence bands at the four inversion-invariant momentum points: $\Gamma$ and three $M$ points.  Let $n_{+}^{k_i}$ ($n_{-}^{k_i}$) denote the number of occupied bands with positive (negative) parity eigenvalue at $k_i$. We find that over the totally 24 valence bands considered in the DFT calculation, at $M$ (the three $M$ are equivalent under the sixfold rotation $C_6$), $n^M_+=13$, and $n_-^M=11$; whereas at $\Gamma$, $n^\Gamma_+=11$, and $n_-^\Gamma=13$. This indicates a nontrivial double band inversion, because the system cannot be adiabatically connected to the trivial atomic insulator limit where the parity representations at the inversion-invariant points must be the same. Indeed, later we shall see that when approaching the trivial atomic limit, the DFT calculation gives  $n_\pm^M=n_\pm^\Gamma=12$ for the trivial phase. The consequence of this nontrivial band inversion will be discussed in more detail in Sec~\ref{sec_topo}.

\begin{figure}[t!]
\includegraphics[width=8 cm]{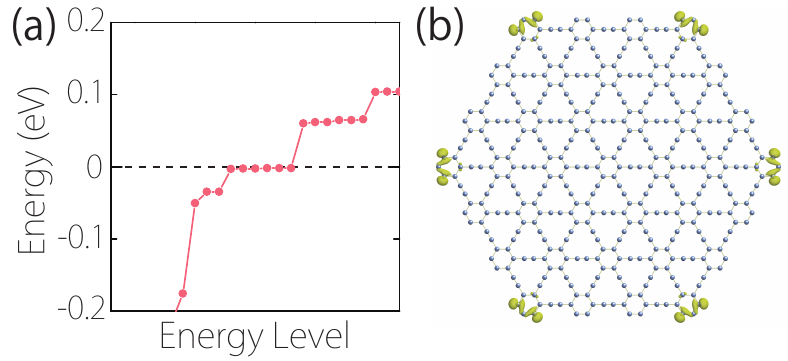}
\caption{Corner states in graphyne. (a) shows the energy spectrum of the hexagonal-shaped graphyne nanodisk in (b). The energy levels are plotted in ascending order. The charge distribution of the six zero-energy modes are illustrated in (b), which are localized at corners.}
\label{zeromode}
\end{figure}

\section{Edge spectra and corner states}\label{sec_edge}

A trivial insulator is usually not expected to have edge states appearing inside the bulk gap. For graphyne, in contrary, we find the existence of a pair of in-gap edge bands for generic edges. For example, in Figs.~\ref{edge}(a) and \ref{edge}(b), we plot the calculated edge spectra corresponding to the zigzag edge and the armchair edge which are illustrated in Figs.~\ref{edge}(c) and \ref{edge}(d). One can observe that each edge has a pair of edge bands (the bright curves in the figures) inside the bulk band gap. Notably, these edge bands are \emph{not} gapless, which is consistent with the bulk not being a conventional TI.
The edge gaps, i.e., the gaps for the edge bands, are 44.0 and 106.4  meV for the zigzag and the armchair edges. In the next section, we shall see that the presence of edge bands is not accidental. It is actually dictated by the bulk band topology.

Next, we explore the existence of corner states, which is considered as the hallmark of a 2D SOTI.
The typical approach is to study the spectrum for a 0D geometry, namely, a graphyne nanodisk. For concrete calculations, we take the nanodisk to be of the hexagonal shape, as shown in Fig.~\ref{zeromode}(b). Later, we will argue that the essential results do not depend on the specific shape.
The obtained discrete energy spectrum for the disk is plotted in Fig.~\ref{zeromode}(a). Notably, there are six zero-energy states in the middle of the spectrum (inside the bulk band gap). By checking their spatial distribution, we confirm that these six states are localized at the six corners of the hexagon, as shown in Fig.~\ref{zeromode}(b). Thus, they indeed correspond to the corner states. For this disk geometry, the degeneracy of these corner states are enforced by the $C_6$ symmetry. It follows that in the ground state, exactly three of the six states are occupied, and the low-energy electronic excitation of the system is gapless and located at the corners of the disk.

\section{Topological origin}\label{sec_topo}

To clarify the topological origin of the peculiar edge spectra and the corner states, we shall proceed in the following steps.
First, we find that the bulk bands of graphyne feature a nontrivial topological invariant, the real Chern number $\nu_R$, so graphyne is an example of a 2D RCI. Then, we show the nontrivial $\nu_R$ of a RCI requires the existence of edge bands and furthermore dictates that the edges are described by the 1D gapped Dirac theory, which has a $\mathbb{Z}_2$ topological classification. Then, a topologically protected 0D state must exist as the domain-wall mode where two edges belonging to distinct classes meet. This is the origin of the corner states found in DFT calculations.

\subsection{Real Chern number}

A crystalline band structure is rich in its symmetry, and in our case the bands of graphyne in the first BZ preserve $D_{6h}$ symmetry together with $\mathcal{T}$. Typically, a subset of symmetries are sufficient for restricting the band topology and the boundary states, and the choice of the subset is often not unique. Here, we shall focus on the combined symmetry $\mathcal{PT}$ of the system.

The $\mathcal{PT}$ symmetry operates locally in the BZ, i.e., any $k$ point is invariant under $\mathcal{PT}$. For a spinless system, $\mathcal{PT}$ satisfies the fundamental relation $(\mathcal{PT})^2=1$, and up to a possible unitary transformation, it can always be represented by
$\mathcal{PT}=\mathcal{K}$, where $\mathcal{K}$ is the complex conjugation. This means that the $\mathcal{PT}$ symmetry requires a spinless system to be described by a \emph{real} band theory, where all the Bloch eigenstates $u_{n}(\bm{k})$ are required to be real, i.e., $u^*_{n}(\bm{k})=u_{n}(\bm{k})$~\cite{ZhaoYX_Schhyder_PRL2016}. Then, for an $M$-band theory with $N$ ($<M$) valence bands, the valence space at each $\bm{k}$ takes its value in the Grassmannian~\cite{RealCN_YXZhao}
\begin{equation}
  \text{Gr}(N,\mathbb{R}^{M})\cong {O(M)}/[{O(N)\times O(M-N)}].
\end{equation}
From the mathematical results of homotopy groups,
\begin{equation}\label{N2}
  \pi_2(\text{Gr}(2,\mathbb{R}^M))=\mathbb{Z},
\end{equation}
and
\begin{equation}\label{Z2}
  \pi_2(\text{Gr}(N,\mathbb{R}^M))=\mathbb{Z}_2,
\end{equation}
for $N\ge 3$ and $M\ge 2N$. We see that there exist strong $\mathcal{PT}$-invariant topological insulators in 2D, characterized by the topological invariant corresponding to the above homotopy class. This invariant is the real Chern number $\nu_R$ (also known as the second Stiefel-Whitney number), which is analogous to the usual Chern number, but characterizes \emph{real} Berry bundles over the BZ rather than complex ones~\cite{RealCN_YXZhao,BJYang_CPB}.

It should be noted that for a 2D real system, the bulk may also possess 1D topological invariants, corresponding to the usual quantized Zak phases for the valence bands~\cite{Zak_PRL1989}.
In fact, a well-defined $\nu_R$ on a 2D BZ requires all Zak phases to vanish. This condition has been confirmed for graphyne.

In the special case of only two valence bands with $N=2$, the topological invariant can be $\mathbb{Z}$-valued according to Eq.~(\ref{N2}), which corresponds to the Euler characteristic class for a 2D real vector bundle over the BZ~\cite{milnor2016characteristic}. However, the Euler class is not stable in $K$-theory, namely, adding any trivial valence band can collapse the $\mathbb{Z}$ classification into $\mathbb{Z}_2$, as seen from Eq.~(\ref{Z2}), which corresponds to the parity of the number. An explicit formula for evaluating $\nu_R$ in the $N=2$ case has been presented in Ref~\cite{RealCN_YXZhao}.

\begin{figure}
  \includegraphics[width=8 cm]{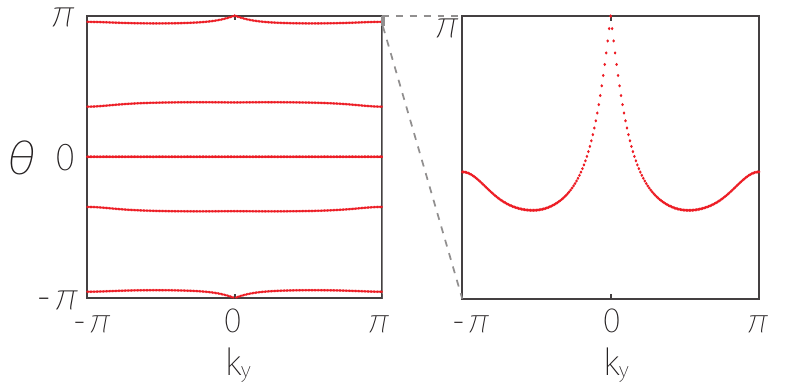}
  \caption{ Wilson loop spectrum for graphyne. The right panel shows the enlarged view near $\theta=\pi$. The calculation is based on a Wannier model including 12 low-energy bands (six occupied and six unoccupied). There are two curves degenerate at $\theta=0$. The spectrum exhibits one crossing point with $\theta = \pi$, which indicates $\nu_R=1$. }
\label{wilson}
\end{figure}

In practice, especially for real materials, $\nu_R$ can be computed by using the Wilson-loop method~\cite{JBYang_PRL2018}, in parallel to the well-known Wilson-loop method for TIs and Chern insulators~\cite{YuRui_PRB2011}. In this method, one calculates the Wilson loop, which is an $N\times N$ matrix, along a particular direction, e.g., along the $k_x$ direction for a fixed $k_y\in(-\pi,\pi]$ in the 2D BZ,
\begin{equation}\label{WL}
  W(k_y)=P\exp\left[-i\int_{C_{k_y}}\mathcal{A}(\bm k)dk_x\right]=\prod_{i=0}^{N_x-1}F_{i,i+1}(k_y),
\end{equation}
where $P$ denotes path ordering, $C_{k_y}$ is the contour at fixed $k_y$, $\mathcal{A}(\bm k)$ is the non-Abelian Berry connection for the valence bands. In the second step of (\ref{WL}), the contour is sampled by discrete $k$ points $(k_{x,i},k_y)$ with a spacing of $2\pi/N_x$, $N_x$ is the number of real-space unit cells along $x$, and in practice the result usually converges quickly with a sufficiently large $N_x$, $F_{i,i+1}$ is the $N\times N$ overlap matrix with elements
\begin{equation}
  F_{i,i+1}^{mn}(k_y)=\langle u_{m}(k_{x,i},k_y)|u_{n}(k_{x,i+1},k_y)\rangle,
\end{equation}
with $m,n=1,2,\cdots,N$, and $F_{N_x-1,N_x}\equiv F_{N_x-1,0}$.
The topological information is encoded in the $N$ phase factors $\theta_m(k_y)\in(-\pi,\pi]$ of the $N$ eigenvalues $\lambda_m(k_y)$ of $W(k_y)$:
\begin{equation}\label{theta}
  \theta_m(k_y)=\text{Im}[\log\lambda_m(k_y)].
\end{equation}
The computation will result in $N$ curves in the $\theta$-$k_y$ diagram (the Wilson loop spectrum), representing the evolution of Wannier centers, as shown in Fig.~\ref{wilson} for graphyne. From the definition, it is clear that the Wilson loop spectrum must be gauge invariant, because under an arbitrary gauge transformation
$|u_n(k_{x,i},k_y)\rangle\rightarrow \sum_m |u_m(k_{x,i},k_y)\rangle [U_i(k_y)]_{mn}$, the Wilson loop operator transforms as $W(k_y)\rightarrow U^\dagger_0(k_y)W(k_y)U_0(k_y)$. It also follows that in evaluating the Wilson loop spectrum, it is not necessary to impose the real gauge (in which all the eigenstates are real).

Note that different from the conventional TIs and Chern insulators, the Wilson loop spectra for real phases are mirror symmetric with respect to the $\theta=0$ axis (see Fig.~\ref{wilson}). To understand this, note that although the eigenstates $|u_n\rangle$ in Eq.~(\ref{WL}) are not required to be real, because of the $\mathcal{PT}$ symmetry, there always exists a unitary transformation $\tilde{U}_i$ to convert all eigenstates to be real. Then, we must have the Wilson loop operator  $W(k_y)=\tilde{U}^\dagger_0(k_y)W^{\mathbb{R}}(k_y)\tilde{U}_{0}(k_y)$, where $W^{\mathbb{R}}$ is a purely real orthogonal matrix. It follows that the eigenvalues of $W^{\mathbb{R}}(k_y)$ and $W(k_y)$ must form complex-conjugation pairs, and therefore the spectra flow of $\theta$ are mirror symmetric with respect to $\theta=0$.

Furthermore, the trace  of a specific pair of bands with spectrum $e^{\pm i \theta_a(k_y)}$ corresponds to a closed path in an $SO(2)$ subgroup that also continuously moves inside $SO(N)$ along $k_y$. (The situation resembles a point moving on a ball that is also moving inside a room.) Restricted in the $SO(2)$ subgroup, closed paths are classified by the winding number $\zeta_a$, and the parity of $\zeta_a$ can be read off from the Wilson loop spectrum as the parity of times that $\theta_a$ goes through $\pi$ according to the chosen branch cut in Eq.~(\ref{theta}). The parity of the sum $\zeta=\sum_a\zeta_a$ is just the topological invariant for the $\mathbb{Z}_2$ classification of the closed path of $W(k_y)$ in $SO(N)$ with $N>2$, namely
\begin{equation}
\nu_R=\zeta \mod 2.
\end{equation}
Note that if $N$ is odd, there is always a flat line with $\theta=0$ in the spectrum, which does not contribute to the topological invariant.
For graphyne, one observes from Fig.~\ref{wilson} that there is a single crossing, hence $\nu_R=1$. This demonstrates that graphyne is a RCI with a nontrivial real Chern number.

As we have mentioned, the definition of $\nu_R$ here is enabled by the combined $\mathcal{PT}$ symmetry. However, when the $\mathcal{P}$ symmetry also exists, there is a shortcut to obtain $\nu_R$ by using the parity eigenvalues of the valence bands at the four inversion-invariant momentum points $\Gamma_i$ $(i=1,\cdots,4)$, namely~\cite{JBYang_PRL2018}
\begin{equation}\label{PE}
(-1)^{\nu_R}=\prod_{i=1}^{4}(-1)^{\left\lfloor (n^{\Gamma_i}_- / 2)\right\rfloor},
\end{equation}
where $\lfloor\cdots\rfloor$ is the floor function.
This approach applies for graphyne. In Sec., we have already obtained $n^{\Gamma_i}_-$ for parity eigenvalues at $\Gamma$ and $M$. Plugging those numbers into (\ref{PE}) immediately gives us $\nu_R=1$, which exactly reproduces the result from the Wilson-loop method.

\subsection{Dirac edge bands}

What is the consequence of the nontrivial bulk invariant $\nu_R$ at the boundary? For a RCI, its bulk Hamiltonian can always be adiabatically deformed into a block diagonal form, consisting of two decoupled blocks (copies) corresponding to a pair of conventional Chern insulators with Chern numbers $\pm 1$, respectively.
It follows that when boundaries are opened, if the two copies were still decoupled at boundary, each edge would have a pair of gapless edge bands with opposite chirality, and hence the boundary would be described by a 1D massless Dirac model. However, in fact,
every edge alone does not preserve $\mathcal{P}$, and therefore the $\mathcal{PT}$ symmetry is violated at the edge. As a result, the two copies are no longer decoupled at the edge. This edge-induced coupling will hybridize the two opposite chiral edge bands, and open a gap in the edge spectrum for a generic edge. This argument thus shows that the edge of a RCI, such as graphyne, should be described by a 1D gapped Dirac model.

Let us be more specific. Assume that we have a graphyne sample which preserves the $\mathcal{P}$ and $\mathcal{T}$ symmetries, such as the disk in Fig.~\ref{zeromode}. Consider one generic edge of the sample. The edge model must preserve the $\mathcal{T}$ symmetry and contain two bands according to the argument above. Taking the representation with $\mathcal{T}=I_k\mathcal{K}$ where $I_k$ is the inversion of the momentum $k$ along the edge, we must have
\begin{equation}\label{H0}
  \mathcal{H}_0=vk\sigma_y
\end{equation}
for the edge model without the edge-induced coupling. Here, the Pauli matrix $\sigma_y$ acts in the Hilbert space spanned by the two states at $k=0$. The edge-induced coupling will introduce additional mass terms to the Dirac model (\ref{H0}). Constrained by $\mathcal{T}$, at leading order, we may have two mass terms $m_1\sigma_x$ and $m_2\sigma_z$. As we mentioned in Sec., graphyne also has an approximate chiral symmetry $\mathcal{C}$, which anti-commutes with the Hamiltonian of the system. Since $[\mathcal{C},\mathcal{T}]=0$, we may choose $\mathcal{C}=\sigma_z$ without loss of generality. Then the condition that $\{\mathcal{C},\mathcal{H}\}=0$ suppresses the $m_2\sigma_z$ term. Therefore, the symmetry-constrained edge model should take the form of
\begin{equation}\label{H}
  \mathcal{H}=vk\sigma_y+m_1\sigma_x,
\end{equation}
which is a 1D gapped Dirac model.

Now, consider the opposite edge of the sample which is connected to the edge (\ref{H}) by the $\mathcal{P}$ operation. Since $\mathcal{P}$ must commute with $\mathcal{T}$ and switches the sign of $\mathcal{H}_0$, it should be represented by $\mathcal{P}=I_k$ in this case. Therefore, the opposite edge is described by
\begin{equation}
  \mathcal{P}^{-1}\mathcal{H}\mathcal{P}=-vk\sigma_y+m_1\sigma_x.
\end{equation}
This shows that the mass term must be reversed between a pair of edges connected by $\mathcal{P}$. It also means that for a generic graphyne sample which preserves $\mathcal{P}$ and $\mathcal{T}$, the mass terms of its edges must have both positive and negative signs.


\subsection{Topological corner states}

The previous two subsections have demonstrated that a generic edge of graphyne is a 1D insulator described by the gapped Dirac model (\ref{H}). It is well known that this model has a $\mathbb{Z}_2$ topological classification by the sign of the mass term $\text{sgn}(m_1)$~\cite{ShunQingShen_TI}. Then, the appearance of corner states becomes natural: When two neighboring edges belong to distinct $\mathbb{Z}_2$ classes, \emph{i.e.}, when their masses have opposite signs, a localized 0D corner state must exist at their joint point.
Essentially, the corner here corresponds to a domain wall between two domains with different topology, and the corner state corresponds to the Jackiw-Rebbi topological domain-wall zero-mode~\cite{Jackiw_PRD} (as illustrated in Supplemental Material).

It should be noted that, by definition, \emph{not every} corner of a 2D SOTI has a protected corner state. For example, if the two neighboring edges belong to the same topological class, then there is no guaranteed corner state between them. Nevertheless, as we have shown in the above Sec., when the graphyne sample preserves $\mathcal{P}$, its edges must have mass terms of different signs, corresponding to different topological classes. In other words, domain walls and the associated 0D topological modes must exist somewhere on the sample boundary, \emph{regardless} of the detailed shape and the boundary geometry. In the next section, we shall further argue that once such corner states exist, they will remain robust even under symmetry-breaking perturbations. This is a general feature of SOTIs.

\begin{figure}[t!]
\includegraphics[width=8 cm]{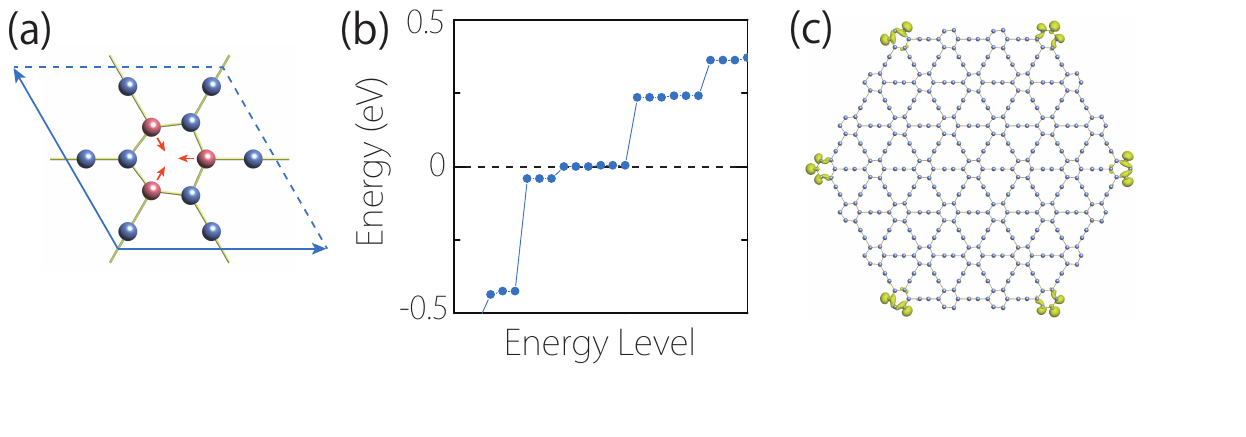}
\caption{Corner states in graphyne under a symmetry breaking perturbation. (a) indicates an artificial distortion that breaks the inversion symmetry. Each red colored atom is displaced from its equilibrium position by 0.14 \AA.  (b) shows the corresponding energy spectrum of the nanodisk geometry in (c). There are still zero-energy states localized at the corners, as illustrated in (c).}
\label{rotate}
\end{figure}

\section{robustness against symmetry breaking}
\label{sec_robust}

In Sec.~\ref{sec_topo}, we have demonstrate topological origin of the corner states in graphyne. In the analysis, we have utilized the crystalline symmetries such as $\mathcal{P}$. However, after revealing the topology of the 1D edge and 0D corner of graphyne, it can be seen that such symmetries are not required to be \emph{exact}~\cite{Langbehn2017,GDY_PRL2019}. This is because
as long as the bulk and edge gaps are not closed, the $\mathbb{Z}_2$ character for the 1D edge cannot be changed by any symmetry-breaking perturbations.
Hence, the corner states, as topological domain-wall modes between the edges, must be robust against the perturbations. In other words, \emph{once} the corner states exist, they cannot be completely annihilated without costing a large amount of energy.

To explicitly demonstrate this point, we break the inversion symmetries by applying an artificial lattice distortion as shown in Fig.~\ref{rotate}(a), and
redo the calculation for the nanodisk geometry. The results are shown in Fig.~\ref{rotate}(b,c). One observes that the zero-energy corner states are indeed maintained. Therefore, the corner states and hence the 2D SOTI phase in graphyne are indeed topologically robust.

\section{Discussion and Conclusion}

Experimentally, the predicted topological corner states in graphyne can be detected by surface-sensitive probes, such as the scanning tunneling spectroscopy (STS) which probes the local density of states (LDOS) with fine energy resolution $\sim 1$ meV and atomic scale spatial resolution. One can compare the STS spectra for tip placed above the center of the nanodisk and for tip above the corner. The existence of corner state should be readily manifested as a sharp peak appearing inside the bulk gap.

In Sec.~\ref{sec_robust}, we have demonstrated that the topological corner states are robust against the symmetry-breaking perturbations, such as lattice distortions.
One can easily see that the argument also applies for impurities, lattice imperfections, geometry variations, and etc.  For example, the 0D boundary states should exist for a generic disk geometry, or even for a disk without sharply defined edges; although for such cases, it could be a challenge to locate them in experimental detection.

The approximate sublattice symmetry $\mathcal{C}$ is beneficial for the detection, because it helps to pin the corner states inside the energy gap. In experiment, the adsorption of impurity atoms or functional groups at edges (especially corners) may locally break the chiral symmetry and shift the corner state energy. This may complicate the experimental detection if the shifted corner-state peak overlaps with the large LDOS from the bulk bands. Thus, the edge contamination needs to be controlled in experimental preparation of the samples.

Finally, we note that graphyne so far has been realized in the form of small fragments~\cite{Diederich1994uk,Bunz1999}. If the sample is too small, the corner states would couple with each other, with their wave functions less localized and their energies pushed away from zero. Nevertheless, our DFT calculation shows that the corner states can still be clearly discerned for a nanodisk with a width as small as $\sim$4 nm. This opens a good opportunity for experimental detection. 
Furthermore, a good news is that the building blocks and cutouts for graphyne have already been synthesized, and the approaches towards extended samples have been proposed and developed. With the additional impetus provided by this work, we expect that extended graphyne can be achieved in the near future.

In conclusion, {we have revealed graphyne as a realistic example of 2D SOTI and 2D RCI. We demonstrate that its Dirac type edge bands and the corner states are dictated by the nontrivial bulk topology characterized by the real Chern number.} We have explicitly evaluated the real Chern number using both the Wilson-loop method and the parity approach. The topological robustness of the corner states has been demonstrated. In addition, we find that the topological phase transition in graphyne is mediated by a 2D Weyl semimetal phase, distinct from previous examples. Our work uncovers a hidden topological character of graphyne. The finding also implies that higher-order TI and RCI may find realizations in a large family of 2D elemental materials. Combined with the many intriguing physical properties already predicted for the graphyne family, the added
topological dimension will promote graphyne as a fascinating platform for both fundamental studies and technological applications.


\begin{appendix}

\renewcommand{\theequation}{A\arabic{equation}}
\setcounter{equation}{0}
\renewcommand{\thefigure}{A\arabic{figure}}
\setcounter{figure}{0}
\renewcommand{\thetable}{A\arabic{table}}
\setcounter{table}{0}

\section{First-Principles Methods}
\label{sec_method}
Our first-principles calculations have been carried out based on the density-functional theory (DFT) as implemented in the Vienna \textit{ab initio} simulation package (VASP)~\cite{Kresse1994,Kresse1996}. The projector augmented wave method~\cite{PAW} was used for treating the ionic potentials.  The generalized gradient approximation (GGA) with the Perdew-Burke-Ernzerhof (PBE)~\cite{PBE} realization was adopted for the exchange-correlation functional.  The plane-wave cutoff energy was set to 500 eV.  The Monkhorst-Pack $k$-point mesh~\cite{PhysRevB.13.5188} of size $11\times11\times 1$ was used for the BZ sampling in the bulk calculations, while the $\Gamma$ point sampling was used for the nanodisk calculations.  Considering the possible underestimation of the band gap by GGA,
the modified Becke-Johnson potential (mBJ)~\cite{mBJ} approach was also used to further check the band structure. The crystal structure was optimized until the forces on the ions were less than 0.01 eV/\AA.  From the DFT results, the maximally localized Wannier functions for the C-$2s$ and C-$2p$ orbitals were constructed, based on which the \emph{ab-initio} tight-binding models were developed to study the edge spectra~\cite{Marzari1997,Souza2001,Wu2017,Green1984,Green1985}.

\section{Band structure with mBJ potential}
\label{sec_mbj}

\begin{figure}[tb]
\includegraphics[width=8 cm]{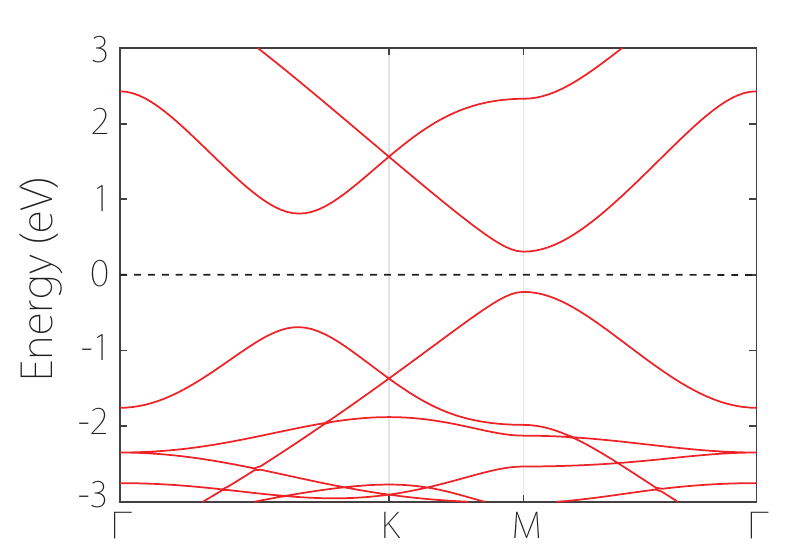}
\caption{Bulk band structure calculated by the mBJ approach.}
\label{mbj}
\end{figure}

The band gap is typically underestimated by GGA. Thus, we further checked the band structure by using the more accurate modified Becke-Johnson potential (mBJ), as shown in Fig.~\ref{mbj}. There is no qualitative difference from the PBE result. The result is still a semiconductor with a direct gap of 0.53 eV at $M$, slightly larger compared to the PBE value of 0.46 eV. The double band inversion feature is also maintained.

\section{Topological corner states}

\begin{figure}[t!]
\includegraphics[width=8 cm]{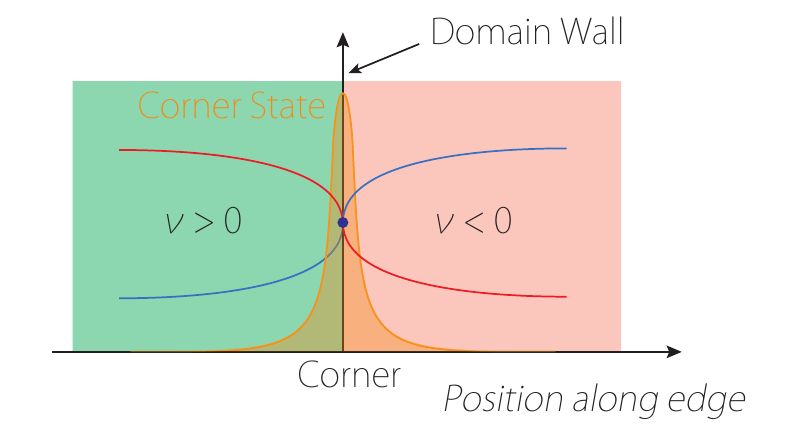}
\caption{Schematic figure showing that two edges with opposite Dirac mass terms form a topological domain wall at the corner, which must host a 0D corner state corresponding to the topological domain-wall mode.}
\label{domainwall}
\end{figure}

The corner state corresponds to a domain wall between two domains with different topology, and the corner state corresponds to the Jackiw-Rebbi topological domain-wall zero-mode~\cite{Jackiw_PRD}, as illustrated in Fig.~\ref{domainwall}.

\section{novel topological phase transition}

\begin{figure}[t!]
\includegraphics[width=8 cm]{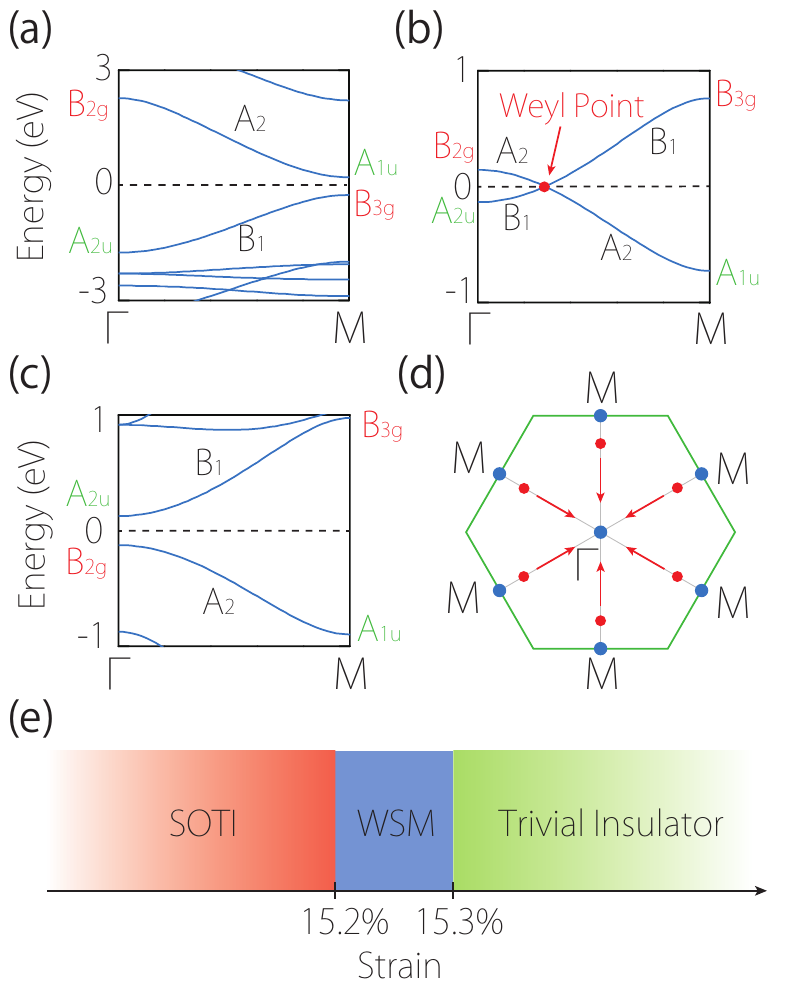}
\caption{Topological and quantum phase transition driven by applied strain. Band structures and their corresponding irreducible representations of (a) SOTI, (b) 2D Weyl semimetal and (c) trivial insulator phases, respectively. (d) shows the evolution of the Weyl points, which emerge at the $M$ points, move towards and finally annihilate at the $\Gamma$ point. (e) Schematic figure showing the topological phase transition.}
\label{strain}
\end{figure}

The SOTI (and RCI) phase is expected to be destroyed when the lattice is expanded towards the atomic insulator limit. This is indeed confirmed by our DFT calculation by applying strain on the graphyne. One finds that above $\sim 15.3\%$ biaxial strain, graphyne will be eventually turned into a trivial insulator.   Interestingly, unlike previously discovered examples, for graphyne, between the SOTI phase and the trivial insulator phase, there exists a 2D Weyl semimetal phase in the range of 15.2\% to 15.3\% strain, as schematically illustrated in Fig.~\ref{strain}(e).

A careful examination of the phase transition shows that at 15.2\% strain, a bandgap closing first happens at the $M$ point, after which the conduction and valence states at $M$ are inverted and a 2D Weyl point is formed on the $\Gamma$-$M$ path, as shown in Fig.~\ref{strain}(b). This Weyl point is protected by the $\mathcal{PT}$ symmetry through the quantized $\pi$ Berry phase for a loop circling around the point.  There are totally six Weyl points in the BZ, and they are moved from $M$ towards $\Gamma$ with increasing strain [see Fig.~\ref{strain}(d)]. When these Weyl points merge at $\Gamma$, they will annihilate, and another band inversion occurs at $\Gamma$, after which the system is eventually turned into a trivial insulator phase [see Fig.~\ref{strain}(c)]. For the trivial insulator, we confirm that $n_\pm^M=n_\pm^\Gamma=12$, and the real Chern number is trivial.

\end{appendix}

\bibliographystyle{apsrev4-1}
\bibliography{GY_ref}

\begin{thebibliography}{61}%
\makeatletter
\providecommand \@ifxundefined [1]{%
 \@ifx{#1\undefined}
}%
\providecommand \@ifnum [1]{%
 \ifnum #1\expandafter \@firstoftwo
 \else \expandafter \@secondoftwo
 \fi
}%
\providecommand \@ifx [1]{%
 \ifx #1\expandafter \@firstoftwo
 \else \expandafter \@secondoftwo
 \fi
}%
\providecommand \natexlab [1]{#1}%
\providecommand \enquote  [1]{``#1''}%
\providecommand \bibnamefont  [1]{#1}%
\providecommand \bibfnamefont [1]{#1}%
\providecommand \citenamefont [1]{#1}%
\providecommand \href@noop [0]{\@secondoftwo}%
\providecommand \href [0]{\begingroup \@sanitize@url \@href}%
\providecommand \@href[1]{\@@startlink{#1}\@@href}%
\providecommand \@@href[1]{\endgroup#1\@@endlink}%
\providecommand \@sanitize@url [0]{\catcode `\\12\catcode `\$12\catcode
  `\&12\catcode `\#12\catcode `\^12\catcode `\_12\catcode `\%12\relax}%
\providecommand \@@startlink[1]{}%
\providecommand \@@endlink[0]{}%
\providecommand \url  [0]{\begingroup\@sanitize@url \@url }%
\providecommand \@url [1]{\endgroup\@href {#1}{\urlprefix }}%
\providecommand \urlprefix  [0]{URL }%
\providecommand \Eprint [0]{\href }%
\providecommand \doibase [0]{http://dx.doi.org/}%
\providecommand \selectlanguage [0]{\@gobble}%
\providecommand \bibinfo  [0]{\@secondoftwo}%
\providecommand \bibfield  [0]{\@secondoftwo}%
\providecommand \translation [1]{[#1]}%
\providecommand \BibitemOpen [0]{}%
\providecommand \bibitemStop [0]{}%
\providecommand \bibitemNoStop [0]{.\EOS\space}%
\providecommand \EOS [0]{\spacefactor3000\relax}%
\providecommand \BibitemShut  [1]{\csname bibitem#1\endcsname}%
\let\auto@bib@innerbib\@empty
\bibitem [{\citenamefont {Baughman}\ \emph {et~al.}(1987)\citenamefont
  {Baughman}, \citenamefont {Eckhardt},\ and\ \citenamefont
  {Kertesz}}]{GDY1987}%
  \BibitemOpen
  \bibfield  {author} {\bibinfo {author} {\bibfnamefont {R.~H.}\ \bibnamefont
  {Baughman}}, \bibinfo {author} {\bibfnamefont {H.}~\bibnamefont {Eckhardt}},
  \ and\ \bibinfo {author} {\bibfnamefont {M.}~\bibnamefont {Kertesz}},\
  }\href@noop {} {\bibfield  {journal} {\bibinfo  {journal} {J. Chem. Phys.}\
  }\textbf {\bibinfo {volume} {87}},\ \bibinfo {pages} {6687} (\bibinfo {year}
  {1987})}\BibitemShut {NoStop}%
\bibitem [{\citenamefont {Diederich}(1994)}]{Diederich1994uk}%
  \BibitemOpen
  \bibfield  {author} {\bibinfo {author} {\bibfnamefont {F.}~\bibnamefont
  {Diederich}},\ }\href@noop {} {\bibfield  {journal} {\bibinfo  {journal}
  {Nature}\ }\textbf {\bibinfo {volume} {369}},\ \bibinfo {pages} {199}
  (\bibinfo {year} {1994})}\BibitemShut {NoStop}%
\bibitem [{\citenamefont {Gao}\ \emph {et~al.}(2019)\citenamefont {Gao},
  \citenamefont {Liu}, \citenamefont {Wang},\ and\ \citenamefont
  {Zhang}}]{ZhangJin_ChemSocRev}%
  \BibitemOpen
  \bibfield  {author} {\bibinfo {author} {\bibfnamefont {X.}~\bibnamefont
  {Gao}}, \bibinfo {author} {\bibfnamefont {H.}~\bibnamefont {Liu}}, \bibinfo
  {author} {\bibfnamefont {D.}~\bibnamefont {Wang}}, \ and\ \bibinfo {author}
  {\bibfnamefont {J.}~\bibnamefont {Zhang}},\ }\href@noop {} {\bibfield
  {journal} {\bibinfo  {journal} {Chem. Soc. Rev.}\ }\textbf {\bibinfo {volume}
  {48}},\ \bibinfo {pages} {908} (\bibinfo {year} {2019})}\BibitemShut
  {NoStop}%
\bibitem [{\citenamefont {Li}\ \emph {et~al.}(2010)\citenamefont {Li},
  \citenamefont {Li}, \citenamefont {Liu}, \citenamefont {Guo}, \citenamefont
  {Li},\ and\ \citenamefont {Zhu}}]{LiYL2010}%
  \BibitemOpen
  \bibfield  {author} {\bibinfo {author} {\bibfnamefont {G.}~\bibnamefont
  {Li}}, \bibinfo {author} {\bibfnamefont {Y.}~\bibnamefont {Li}}, \bibinfo
  {author} {\bibfnamefont {H.}~\bibnamefont {Liu}}, \bibinfo {author}
  {\bibfnamefont {Y.}~\bibnamefont {Guo}}, \bibinfo {author} {\bibfnamefont
  {Y.}~\bibnamefont {Li}}, \ and\ \bibinfo {author} {\bibfnamefont
  {D.}~\bibnamefont {Zhu}},\ }\href@noop {} {\bibfield  {journal} {\bibinfo
  {journal} {Chem. Commun.}\ }\textbf {\bibinfo {volume} {46}},\ \bibinfo
  {pages} {3256} (\bibinfo {year} {2010})}\BibitemShut {NoStop}%
\bibitem [{\citenamefont {H.~F.~Bunz}\ \emph {et~al.}(1999)\citenamefont
  {H.~F.~Bunz}, \citenamefont {Rubin},\ and\ \citenamefont {Tobe}}]{Bunz1999}%
  \BibitemOpen
  \bibfield  {author} {\bibinfo {author} {\bibfnamefont {U.}~\bibnamefont
  {H.~F.~Bunz}}, \bibinfo {author} {\bibfnamefont {Y.}~\bibnamefont {Rubin}}, \
  and\ \bibinfo {author} {\bibfnamefont {Y.}~\bibnamefont {Tobe}},\ }\href@noop
  {} {\bibfield  {journal} {\bibinfo  {journal} {Chem. Soc. Rev.}\ }\textbf
  {\bibinfo {volume} {28}},\ \bibinfo {pages} {107} (\bibinfo {year}
  {1999})}\BibitemShut {NoStop}%
\bibitem [{\citenamefont {Ajori}\ \emph {et~al.}(2013)\citenamefont {Ajori},
  \citenamefont {Ansari},\ and\ \citenamefont {Mirnezhad}}]{Ajori2013}%
  \BibitemOpen
  \bibfield  {author} {\bibinfo {author} {\bibfnamefont {S.}~\bibnamefont
  {Ajori}}, \bibinfo {author} {\bibfnamefont {R.}~\bibnamefont {Ansari}}, \
  and\ \bibinfo {author} {\bibfnamefont {M.}~\bibnamefont {Mirnezhad}},\
  }\href@noop {} {\bibfield  {journal} {\bibinfo  {journal} {Mater. Sci. Eng.
  A}\ }\textbf {\bibinfo {volume} {561}},\ \bibinfo {pages} {34 } (\bibinfo
  {year} {2013})}\BibitemShut {NoStop}%
\bibitem [{\citenamefont {Azizi}\ \emph {et~al.}(2020)\citenamefont {Azizi},
  \citenamefont {Rezaee}, \citenamefont {Hadianfard},\ and\ \citenamefont
  {Dehnou}}]{Babak2020}%
  \BibitemOpen
  \bibfield  {author} {\bibinfo {author} {\bibfnamefont {B.}~\bibnamefont
  {Azizi}}, \bibinfo {author} {\bibfnamefont {S.}~\bibnamefont {Rezaee}},
  \bibinfo {author} {\bibfnamefont {M.~J.}\ \bibnamefont {Hadianfard}}, \ and\
  \bibinfo {author} {\bibfnamefont {K.~H.}\ \bibnamefont {Dehnou}},\
  }\href@noop {} {\bibfield  {journal} {\bibinfo  {journal} {Comput. Mater.
  Sci.}\ }\textbf {\bibinfo {volume} {182}},\ \bibinfo {pages} {109794}
  (\bibinfo {year} {2020})}\BibitemShut {NoStop}%
\bibitem [{\citenamefont {Li}\ \emph {et~al.}(2015)\citenamefont {Li},
  \citenamefont {Smeu}, \citenamefont {Rives}, \citenamefont {Maraval},
  \citenamefont {Chauvin}, \citenamefont {Ratner},\ and\ \citenamefont
  {Borguet}}]{Li2015nc}%
  \BibitemOpen
  \bibfield  {author} {\bibinfo {author} {\bibfnamefont {Z.}~\bibnamefont
  {Li}}, \bibinfo {author} {\bibfnamefont {M.}~\bibnamefont {Smeu}}, \bibinfo
  {author} {\bibfnamefont {A.}~\bibnamefont {Rives}}, \bibinfo {author}
  {\bibfnamefont {V.}~\bibnamefont {Maraval}}, \bibinfo {author} {\bibfnamefont
  {R.}~\bibnamefont {Chauvin}}, \bibinfo {author} {\bibfnamefont {M.~A.}\
  \bibnamefont {Ratner}}, \ and\ \bibinfo {author} {\bibfnamefont
  {E.}~\bibnamefont {Borguet}},\ }\href@noop {} {\bibfield  {journal} {\bibinfo
   {journal} {Nat. Comm.}\ }\textbf {\bibinfo {volume} {6}},\ \bibinfo {pages}
  {6321} (\bibinfo {year} {2015})}\BibitemShut {NoStop}%
\bibitem [{\citenamefont {Chen}\ \emph {et~al.}(2013)\citenamefont {Chen},
  \citenamefont {Xi}, \citenamefont {Wang},\ and\ \citenamefont
  {Shuai}}]{ShuaiZG2013jpcl}%
  \BibitemOpen
  \bibfield  {author} {\bibinfo {author} {\bibfnamefont {J.}~\bibnamefont
  {Chen}}, \bibinfo {author} {\bibfnamefont {J.}~\bibnamefont {Xi}}, \bibinfo
  {author} {\bibfnamefont {D.}~\bibnamefont {Wang}}, \ and\ \bibinfo {author}
  {\bibfnamefont {Z.}~\bibnamefont {Shuai}},\ }\href@noop {} {\bibfield
  {journal} {\bibinfo  {journal} {J. Phys. Chem. Lett.}\ }\textbf {\bibinfo
  {volume} {4}},\ \bibinfo {pages} {1443} (\bibinfo {year} {2013})}\BibitemShut
  {NoStop}%
\bibitem [{\citenamefont {Yang}\ \emph {et~al.}(2013)\citenamefont {Yang},
  \citenamefont {Liu}, \citenamefont {Wen}, \citenamefont {Tang}, \citenamefont
  {Zhao}, \citenamefont {Li},\ and\ \citenamefont {Wang}}]{Yang2013cm}%
  \BibitemOpen
  \bibfield  {author} {\bibinfo {author} {\bibfnamefont {N.}~\bibnamefont
  {Yang}}, \bibinfo {author} {\bibfnamefont {Y.}~\bibnamefont {Liu}}, \bibinfo
  {author} {\bibfnamefont {H.}~\bibnamefont {Wen}}, \bibinfo {author}
  {\bibfnamefont {Z.}~\bibnamefont {Tang}}, \bibinfo {author} {\bibfnamefont
  {H.}~\bibnamefont {Zhao}}, \bibinfo {author} {\bibfnamefont {Y.}~\bibnamefont
  {Li}}, \ and\ \bibinfo {author} {\bibfnamefont {D.}~\bibnamefont {Wang}},\
  }\href@noop {} {\bibfield  {journal} {\bibinfo  {journal} {ACS Nano}\
  }\textbf {\bibinfo {volume} {7}},\ \bibinfo {pages} {1504} (\bibinfo {year}
  {2013})}\BibitemShut {NoStop}%
\bibitem [{\citenamefont {Tang}\ \emph {et~al.}(2014)\citenamefont {Tang},
  \citenamefont {Hessel}, \citenamefont {Wang}, \citenamefont {Yang},
  \citenamefont {Yu}, \citenamefont {Zhao},\ and\ \citenamefont
  {Wang}}]{WangD2014}%
  \BibitemOpen
  \bibfield  {author} {\bibinfo {author} {\bibfnamefont {H.}~\bibnamefont
  {Tang}}, \bibinfo {author} {\bibfnamefont {C.~M.}\ \bibnamefont {Hessel}},
  \bibinfo {author} {\bibfnamefont {J.}~\bibnamefont {Wang}}, \bibinfo {author}
  {\bibfnamefont {N.}~\bibnamefont {Yang}}, \bibinfo {author} {\bibfnamefont
  {R.}~\bibnamefont {Yu}}, \bibinfo {author} {\bibfnamefont {H.}~\bibnamefont
  {Zhao}}, \ and\ \bibinfo {author} {\bibfnamefont {D.}~\bibnamefont {Wang}},\
  }\href@noop {} {\bibfield  {journal} {\bibinfo  {journal} {Chem. Soc. Rev.}\
  }\textbf {\bibinfo {volume} {43}},\ \bibinfo {pages} {4281} (\bibinfo {year}
  {2014})}\BibitemShut {NoStop}%
\bibitem [{\citenamefont {Xue}\ \emph {et~al.}(2018)\citenamefont {Xue},
  \citenamefont {Huang}, \citenamefont {Yi}, \citenamefont {Guo}, \citenamefont
  {Zuo}, \citenamefont {Li}, \citenamefont {Jia}, \citenamefont {Liu},\ and\
  \citenamefont {Li}}]{Xue2018vq}%
  \BibitemOpen
  \bibfield  {author} {\bibinfo {author} {\bibfnamefont {Y.}~\bibnamefont
  {Xue}}, \bibinfo {author} {\bibfnamefont {B.}~\bibnamefont {Huang}}, \bibinfo
  {author} {\bibfnamefont {Y.}~\bibnamefont {Yi}}, \bibinfo {author}
  {\bibfnamefont {Y.}~\bibnamefont {Guo}}, \bibinfo {author} {\bibfnamefont
  {Z.}~\bibnamefont {Zuo}}, \bibinfo {author} {\bibfnamefont {Y.}~\bibnamefont
  {Li}}, \bibinfo {author} {\bibfnamefont {Z.}~\bibnamefont {Jia}}, \bibinfo
  {author} {\bibfnamefont {H.}~\bibnamefont {Liu}}, \ and\ \bibinfo {author}
  {\bibfnamefont {Y.}~\bibnamefont {Li}},\ }\href@noop {} {\bibfield  {journal}
  {\bibinfo  {journal} {Nat. Comm.}\ }\textbf {\bibinfo {volume} {9}},\
  \bibinfo {pages} {1460} (\bibinfo {year} {2018})}\BibitemShut {NoStop}%
\bibitem [{\citenamefont {Zhao}\ \emph {et~al.}(2015)\citenamefont {Zhao},
  \citenamefont {Wei}, \citenamefont {Zhou}, \citenamefont {Shi},\ and\
  \citenamefont {Zhou}}]{Zhao2015ComMatSci}%
  \BibitemOpen
  \bibfield  {author} {\bibinfo {author} {\bibfnamefont {H.}~\bibnamefont
  {Zhao}}, \bibinfo {author} {\bibfnamefont {D.}~\bibnamefont {Wei}}, \bibinfo
  {author} {\bibfnamefont {L.}~\bibnamefont {Zhou}}, \bibinfo {author}
  {\bibfnamefont {H.}~\bibnamefont {Shi}}, \ and\ \bibinfo {author}
  {\bibfnamefont {X.}~\bibnamefont {Zhou}},\ }\href@noop {} {\bibfield
  {journal} {\bibinfo  {journal} {Comput. Mater. Sci.}\ }\textbf {\bibinfo
  {volume} {106}},\ \bibinfo {pages} {69 } (\bibinfo {year}
  {2015})}\BibitemShut {NoStop}%
\bibitem [{\citenamefont {Ouyang}\ \emph {et~al.}(2012)\citenamefont {Ouyang},
  \citenamefont {Chen}, \citenamefont {Liu}, \citenamefont {Xie}, \citenamefont
  {Wei},\ and\ \citenamefont {Zhong}}]{Tao_PhysRevB2012}%
  \BibitemOpen
  \bibfield  {author} {\bibinfo {author} {\bibfnamefont {T.}~\bibnamefont
  {Ouyang}}, \bibinfo {author} {\bibfnamefont {Y.}~\bibnamefont {Chen}},
  \bibinfo {author} {\bibfnamefont {L.-M.}\ \bibnamefont {Liu}}, \bibinfo
  {author} {\bibfnamefont {Y.}~\bibnamefont {Xie}}, \bibinfo {author}
  {\bibfnamefont {X.}~\bibnamefont {Wei}}, \ and\ \bibinfo {author}
  {\bibfnamefont {J.}~\bibnamefont {Zhong}},\ }\href@noop {} {\bibfield
  {journal} {\bibinfo  {journal} {Phys. Rev. B}\ }\textbf {\bibinfo {volume}
  {85}},\ \bibinfo {pages} {235436} (\bibinfo {year} {2012})}\BibitemShut
  {NoStop}%
\bibitem [{\citenamefont {Solis}\ \emph {et~al.}(2019)\citenamefont {Solis},
  \citenamefont {D.~Borges}, \citenamefont {Woellner},\ and\ \citenamefont
  {Galvão}}]{Solis_ACS2019}%
  \BibitemOpen
  \bibfield  {author} {\bibinfo {author} {\bibfnamefont {D.~A.}\ \bibnamefont
  {Solis}}, \bibinfo {author} {\bibfnamefont {D.}~\bibnamefont {D.~Borges}},
  \bibinfo {author} {\bibfnamefont {C.~F.}\ \bibnamefont {Woellner}}, \ and\
  \bibinfo {author} {\bibfnamefont {D.~S.}\ \bibnamefont {Galvão}},\
  }\href@noop {} {\bibfield  {journal} {\bibinfo  {journal} {ACS Appl. Mater.
  Interfaces}\ }\textbf {\bibinfo {volume} {11}},\ \bibinfo {pages} {2670}
  (\bibinfo {year} {2019})}\BibitemShut {NoStop}%
\bibitem [{\citenamefont {Malko}\ \emph {et~al.}(2012)\citenamefont {Malko},
  \citenamefont {Neiss}, \citenamefont {Vi\~nes},\ and\ \citenamefont
  {G\"orling}}]{Malko_PRL}%
  \BibitemOpen
  \bibfield  {author} {\bibinfo {author} {\bibfnamefont {D.}~\bibnamefont
  {Malko}}, \bibinfo {author} {\bibfnamefont {C.}~\bibnamefont {Neiss}},
  \bibinfo {author} {\bibfnamefont {F.}~\bibnamefont {Vi\~nes}}, \ and\
  \bibinfo {author} {\bibfnamefont {A.}~\bibnamefont {G\"orling}},\ }\href@noop
  {} {\bibfield  {journal} {\bibinfo  {journal} {Phys. Rev. Lett.}\ }\textbf
  {\bibinfo {volume} {108}},\ \bibinfo {pages} {086804} (\bibinfo {year}
  {2012})}\BibitemShut {NoStop}%
\bibitem [{\citenamefont {Wang}\ \emph {et~al.}(2015)\citenamefont {Wang},
  \citenamefont {Deng}, \citenamefont {Liu},\ and\ \citenamefont
  {Liu}}]{NatSciRev2015}%
  \BibitemOpen
  \bibfield  {author} {\bibinfo {author} {\bibfnamefont {J.}~\bibnamefont
  {Wang}}, \bibinfo {author} {\bibfnamefont {S.}~\bibnamefont {Deng}}, \bibinfo
  {author} {\bibfnamefont {Z.}~\bibnamefont {Liu}}, \ and\ \bibinfo {author}
  {\bibfnamefont {Z.}~\bibnamefont {Liu}},\ }\href@noop {} {\bibfield
  {journal} {\bibinfo  {journal} {Nat. Sci. Rev.}\ }\textbf {\bibinfo {volume}
  {2}},\ \bibinfo {pages} {22} (\bibinfo {year} {2015})}\BibitemShut {NoStop}%
\bibitem [{\citenamefont {Min}\ \emph {et~al.}(2006)\citenamefont {Min},
  \citenamefont {Hill}, \citenamefont {Sinitsyn}, \citenamefont {Sahu},
  \citenamefont {Kleinman},\ and\ \citenamefont {MacDonald}}]{Hongki_PRB}%
  \BibitemOpen
  \bibfield  {author} {\bibinfo {author} {\bibfnamefont {H.}~\bibnamefont
  {Min}}, \bibinfo {author} {\bibfnamefont {J.~E.}\ \bibnamefont {Hill}},
  \bibinfo {author} {\bibfnamefont {N.~A.}\ \bibnamefont {Sinitsyn}}, \bibinfo
  {author} {\bibfnamefont {B.~R.}\ \bibnamefont {Sahu}}, \bibinfo {author}
  {\bibfnamefont {L.}~\bibnamefont {Kleinman}}, \ and\ \bibinfo {author}
  {\bibfnamefont {A.~H.}\ \bibnamefont {MacDonald}},\ }\href@noop {} {\bibfield
   {journal} {\bibinfo  {journal} {Phys. Rev. B}\ }\textbf {\bibinfo {volume}
  {74}},\ \bibinfo {pages} {165310} (\bibinfo {year} {2006})}\BibitemShut
  {NoStop}%
\bibitem [{\citenamefont {Yao}\ \emph {et~al.}(2007)\citenamefont {Yao},
  \citenamefont {Ye}, \citenamefont {Qi}, \citenamefont {Zhang},\ and\
  \citenamefont {Fang}}]{YuguiYao_PRB}%
  \BibitemOpen
  \bibfield  {author} {\bibinfo {author} {\bibfnamefont {Y.}~\bibnamefont
  {Yao}}, \bibinfo {author} {\bibfnamefont {F.}~\bibnamefont {Ye}}, \bibinfo
  {author} {\bibfnamefont {X.-L.}\ \bibnamefont {Qi}}, \bibinfo {author}
  {\bibfnamefont {S.-C.}\ \bibnamefont {Zhang}}, \ and\ \bibinfo {author}
  {\bibfnamefont {Z.}~\bibnamefont {Fang}},\ }\href@noop {} {\bibfield
  {journal} {\bibinfo  {journal} {Phys. Rev. B}\ }\textbf {\bibinfo {volume}
  {75}},\ \bibinfo {pages} {041401} (\bibinfo {year} {2007})}\BibitemShut
  {NoStop}%
\bibitem [{\citenamefont {Narita}\ \emph {et~al.}(1998)\citenamefont {Narita},
  \citenamefont {Nagai}, \citenamefont {Suzuki},\ and\ \citenamefont
  {Nakao}}]{Narita1998}%
  \BibitemOpen
  \bibfield  {author} {\bibinfo {author} {\bibfnamefont {N.}~\bibnamefont
  {Narita}}, \bibinfo {author} {\bibfnamefont {S.}~\bibnamefont {Nagai}},
  \bibinfo {author} {\bibfnamefont {S.}~\bibnamefont {Suzuki}}, \ and\ \bibinfo
  {author} {\bibfnamefont {K.}~\bibnamefont {Nakao}},\ }\href@noop {}
  {\bibfield  {journal} {\bibinfo  {journal} {Phys. Rev. B}\ }\textbf {\bibinfo
  {volume} {58}},\ \bibinfo {pages} {11009} (\bibinfo {year}
  {1998})}\BibitemShut {NoStop}%
\bibitem [{\citenamefont {Zhang}\ \emph {et~al.}(2013)\citenamefont {Zhang},
  \citenamefont {Kane},\ and\ \citenamefont {Mele}}]{ZhangFan_PRL2013}%
  \BibitemOpen
  \bibfield  {author} {\bibinfo {author} {\bibfnamefont {F.}~\bibnamefont
  {Zhang}}, \bibinfo {author} {\bibfnamefont {C.~L.}\ \bibnamefont {Kane}}, \
  and\ \bibinfo {author} {\bibfnamefont {E.~J.}\ \bibnamefont {Mele}},\
  }\href@noop {} {\bibfield  {journal} {\bibinfo  {journal} {Phys. Rev. Lett.}\
  }\textbf {\bibinfo {volume} {110}},\ \bibinfo {pages} {046404} (\bibinfo
  {year} {2013})}\BibitemShut {NoStop}%
\bibitem [{\citenamefont {Benalcazar}\ \emph {et~al.}(2017)\citenamefont
  {Benalcazar}, \citenamefont {Bernevig},\ and\ \citenamefont
  {Hughes}}]{Hughes2017}%
  \BibitemOpen
  \bibfield  {author} {\bibinfo {author} {\bibfnamefont {W.~A.}\ \bibnamefont
  {Benalcazar}}, \bibinfo {author} {\bibfnamefont {B.~A.}\ \bibnamefont
  {Bernevig}}, \ and\ \bibinfo {author} {\bibfnamefont {T.~L.}\ \bibnamefont
  {Hughes}},\ }\href@noop {} {\bibfield  {journal} {\bibinfo  {journal}
  {Science}\ }\textbf {\bibinfo {volume} {357}},\ \bibinfo {pages} {61}
  (\bibinfo {year} {2017})}\BibitemShut {NoStop}%
\bibitem [{\citenamefont {Langbehn}\ \emph {et~al.}(2017)\citenamefont
  {Langbehn}, \citenamefont {Peng}, \citenamefont {Trifunovic}, \citenamefont
  {von Oppen},\ and\ \citenamefont {Brouwer}}]{Langbehn2017}%
  \BibitemOpen
  \bibfield  {author} {\bibinfo {author} {\bibfnamefont {J.}~\bibnamefont
  {Langbehn}}, \bibinfo {author} {\bibfnamefont {Y.}~\bibnamefont {Peng}},
  \bibinfo {author} {\bibfnamefont {L.}~\bibnamefont {Trifunovic}}, \bibinfo
  {author} {\bibfnamefont {F.}~\bibnamefont {von Oppen}}, \ and\ \bibinfo
  {author} {\bibfnamefont {P.~W.}\ \bibnamefont {Brouwer}},\ }\href@noop {}
  {\bibfield  {journal} {\bibinfo  {journal} {Phys. Rev. Lett.}\ }\textbf
  {\bibinfo {volume} {119}},\ \bibinfo {pages} {246401} (\bibinfo {year}
  {2017})}\BibitemShut {NoStop}%
\bibitem [{\citenamefont {Song}\ \emph {et~al.}(2017)\citenamefont {Song},
  \citenamefont {Fang},\ and\ \citenamefont {Fang}}]{SongZD2017}%
  \BibitemOpen
  \bibfield  {author} {\bibinfo {author} {\bibfnamefont {Z.}~\bibnamefont
  {Song}}, \bibinfo {author} {\bibfnamefont {Z.}~\bibnamefont {Fang}}, \ and\
  \bibinfo {author} {\bibfnamefont {C.}~\bibnamefont {Fang}},\ }\href@noop {}
  {\bibfield  {journal} {\bibinfo  {journal} {Phys. Rev. Lett.}\ }\textbf
  {\bibinfo {volume} {119}},\ \bibinfo {pages} {246402} (\bibinfo {year}
  {2017})}\BibitemShut {NoStop}%
\bibitem [{\citenamefont {Schindler}\ \emph
  {et~al.}(2018{\natexlab{a}})\citenamefont {Schindler}, \citenamefont {Cook},
  \citenamefont {Vergniory}, \citenamefont {Wang}, \citenamefont {Parkin},
  \citenamefont {Bernevig},\ and\ \citenamefont {Neupert}}]{Schindler2018SA}%
  \BibitemOpen
  \bibfield  {author} {\bibinfo {author} {\bibfnamefont {F.}~\bibnamefont
  {Schindler}}, \bibinfo {author} {\bibfnamefont {A.~M.}\ \bibnamefont {Cook}},
  \bibinfo {author} {\bibfnamefont {M.~G.}\ \bibnamefont {Vergniory}}, \bibinfo
  {author} {\bibfnamefont {Z.}~\bibnamefont {Wang}}, \bibinfo {author}
  {\bibfnamefont {S.~S.~P.}\ \bibnamefont {Parkin}}, \bibinfo {author}
  {\bibfnamefont {B.~A.}\ \bibnamefont {Bernevig}}, \ and\ \bibinfo {author}
  {\bibfnamefont {T.}~\bibnamefont {Neupert}},\ }\href@noop {} {\bibfield
  {journal} {\bibinfo  {journal} {Sci. Adv.}\ }\textbf {\bibinfo {volume} {4}}
  (\bibinfo {year} {2018}{\natexlab{a}})}\BibitemShut {NoStop}%
\bibitem [{\citenamefont {Kruthoff}\ \emph {et~al.}(2017)\citenamefont
  {Kruthoff}, \citenamefont {de~Boer}, \citenamefont {van Wezel}, \citenamefont
  {Kane},\ and\ \citenamefont {Slager}}]{Kruthoff2017}%
  \BibitemOpen
  \bibfield  {author} {\bibinfo {author} {\bibfnamefont {J.}~\bibnamefont
  {Kruthoff}}, \bibinfo {author} {\bibfnamefont {J.}~\bibnamefont {de~Boer}},
  \bibinfo {author} {\bibfnamefont {J.}~\bibnamefont {van Wezel}}, \bibinfo
  {author} {\bibfnamefont {C.~L.}\ \bibnamefont {Kane}}, \ and\ \bibinfo
  {author} {\bibfnamefont {R.-J.}\ \bibnamefont {Slager}},\ }\href@noop {}
  {\bibfield  {journal} {\bibinfo  {journal} {Phys. Rev. X}\ }\textbf {\bibinfo
  {volume} {7}},\ \bibinfo {pages} {041069} (\bibinfo {year}
  {2017})}\BibitemShut {NoStop}%
\bibitem [{\citenamefont {Ezawa}(2018)}]{Ezawa2018L}%
  \BibitemOpen
  \bibfield  {author} {\bibinfo {author} {\bibfnamefont {M.}~\bibnamefont
  {Ezawa}},\ }\href@noop {} {\bibfield  {journal} {\bibinfo  {journal} {Phys.
  Rev. Lett.}\ }\textbf {\bibinfo {volume} {120}},\ \bibinfo {pages} {026801}
  (\bibinfo {year} {2018})}\BibitemShut {NoStop}%
\bibitem [{\citenamefont {Imhof}\ \emph {et~al.}(2018)\citenamefont {Imhof},
  \citenamefont {Berger}, \citenamefont {Bayer}, \citenamefont {Brehm},
  \citenamefont {Molenkamp}, \citenamefont {Kiessling}, \citenamefont
  {Schindler}, \citenamefont {Lee}, \citenamefont {Greiter}, \citenamefont
  {Neupert},\ and\ \citenamefont {Thomale}}]{Imhof2018wj}%
  \BibitemOpen
  \bibfield  {author} {\bibinfo {author} {\bibfnamefont {S.}~\bibnamefont
  {Imhof}}, \bibinfo {author} {\bibfnamefont {C.}~\bibnamefont {Berger}},
  \bibinfo {author} {\bibfnamefont {F.}~\bibnamefont {Bayer}}, \bibinfo
  {author} {\bibfnamefont {J.}~\bibnamefont {Brehm}}, \bibinfo {author}
  {\bibfnamefont {L.~W.}\ \bibnamefont {Molenkamp}}, \bibinfo {author}
  {\bibfnamefont {T.}~\bibnamefont {Kiessling}}, \bibinfo {author}
  {\bibfnamefont {F.}~\bibnamefont {Schindler}}, \bibinfo {author}
  {\bibfnamefont {C.~H.}\ \bibnamefont {Lee}}, \bibinfo {author} {\bibfnamefont
  {M.}~\bibnamefont {Greiter}}, \bibinfo {author} {\bibfnamefont
  {T.}~\bibnamefont {Neupert}}, \ and\ \bibinfo {author} {\bibfnamefont
  {R.}~\bibnamefont {Thomale}},\ }\href@noop {} {\bibfield  {journal} {\bibinfo
   {journal} {Nat. Phys.}\ }\textbf {\bibinfo {volume} {14}},\ \bibinfo {pages}
  {925} (\bibinfo {year} {2018})}\BibitemShut {NoStop}%
\bibitem [{\citenamefont {Serra-Garcia}\ \emph {et~al.}(2018)\citenamefont
  {Serra-Garcia}, \citenamefont {Peri}, \citenamefont {S{\"u}sstrunk},
  \citenamefont {Bilal}, \citenamefont {Larsen}, \citenamefont {Villanueva},\
  and\ \citenamefont {Huber}}]{SerraGarcia2018}%
  \BibitemOpen
  \bibfield  {author} {\bibinfo {author} {\bibfnamefont {M.}~\bibnamefont
  {Serra-Garcia}}, \bibinfo {author} {\bibfnamefont {V.}~\bibnamefont {Peri}},
  \bibinfo {author} {\bibfnamefont {R.}~\bibnamefont {S{\"u}sstrunk}}, \bibinfo
  {author} {\bibfnamefont {O.~R.}\ \bibnamefont {Bilal}}, \bibinfo {author}
  {\bibfnamefont {T.}~\bibnamefont {Larsen}}, \bibinfo {author} {\bibfnamefont
  {L.~G.}\ \bibnamefont {Villanueva}}, \ and\ \bibinfo {author} {\bibfnamefont
  {S.~D.}\ \bibnamefont {Huber}},\ }\href@noop {} {\bibfield  {journal}
  {\bibinfo  {journal} {Nature}\ }\textbf {\bibinfo {volume} {555}},\ \bibinfo
  {pages} {342} (\bibinfo {year} {2018})}\BibitemShut {NoStop}%
\bibitem [{\citenamefont {Peterson}\ \emph {et~al.}(2018)\citenamefont
  {Peterson}, \citenamefont {Benalcazar}, \citenamefont {Hughes},\ and\
  \citenamefont {Bahl}}]{Peterson2018}%
  \BibitemOpen
  \bibfield  {author} {\bibinfo {author} {\bibfnamefont {C.~W.}\ \bibnamefont
  {Peterson}}, \bibinfo {author} {\bibfnamefont {W.~A.}\ \bibnamefont
  {Benalcazar}}, \bibinfo {author} {\bibfnamefont {T.~L.}\ \bibnamefont
  {Hughes}}, \ and\ \bibinfo {author} {\bibfnamefont {G.}~\bibnamefont
  {Bahl}},\ }\href@noop {} {\bibfield  {journal} {\bibinfo  {journal} {Nature}\
  }\textbf {\bibinfo {volume} {555}},\ \bibinfo {pages} {346} (\bibinfo {year}
  {2018})}\BibitemShut {NoStop}%
\bibitem [{\citenamefont {Liu}\ \emph {et~al.}(2020)\citenamefont {Liu},
  \citenamefont {Wang}, \citenamefont {Hu}, \citenamefont {Lin}, \citenamefont
  {Lee},\ and\ \citenamefont {Zhang}}]{Yuhan_arXiv}%
  \BibitemOpen
  \bibfield  {author} {\bibinfo {author} {\bibfnamefont {Y.}~\bibnamefont
  {Liu}}, \bibinfo {author} {\bibfnamefont {Y.}~\bibnamefont {Wang}}, \bibinfo
  {author} {\bibfnamefont {N.~C.}\ \bibnamefont {Hu}}, \bibinfo {author}
  {\bibfnamefont {J.~Y.}\ \bibnamefont {Lin}}, \bibinfo {author} {\bibfnamefont
  {C.~H.}\ \bibnamefont {Lee}}, \ and\ \bibinfo {author} {\bibfnamefont
  {X.}~\bibnamefont {Zhang}},\ }\href@noop {} {\bibfield  {journal} {\bibinfo
  {journal} {Phys. Rev. B}\ }\textbf {\bibinfo {volume} {102}},\ \bibinfo
  {pages} {035142} (\bibinfo {year} {2020})}\BibitemShut {NoStop}%
\bibitem [{\citenamefont {Xue}\ \emph {et~al.}(2019)\citenamefont {Xue},
  \citenamefont {Yang}, \citenamefont {Gao}, \citenamefont {Chong},\ and\
  \citenamefont {Zhang}}]{Xue2019to}%
  \BibitemOpen
  \bibfield  {author} {\bibinfo {author} {\bibfnamefont {H.}~\bibnamefont
  {Xue}}, \bibinfo {author} {\bibfnamefont {Y.}~\bibnamefont {Yang}}, \bibinfo
  {author} {\bibfnamefont {F.}~\bibnamefont {Gao}}, \bibinfo {author}
  {\bibfnamefont {Y.}~\bibnamefont {Chong}}, \ and\ \bibinfo {author}
  {\bibfnamefont {B.}~\bibnamefont {Zhang}},\ }\href@noop {} {\bibfield
  {journal} {\bibinfo  {journal} {Nat. Mater.}\ }\textbf {\bibinfo {volume}
  {18}},\ \bibinfo {pages} {108} (\bibinfo {year} {2019})}\BibitemShut
  {NoStop}%
\bibitem [{\citenamefont {Yang}\ \emph {et~al.}(2020)\citenamefont {Yang},
  \citenamefont {Jia}, \citenamefont {Wu}, \citenamefont {Xiao}, \citenamefont
  {Hang}, \citenamefont {Jiang},\ and\ \citenamefont {Xie}}]{JiangH2020}%
  \BibitemOpen
  \bibfield  {author} {\bibinfo {author} {\bibfnamefont {Y.}~\bibnamefont
  {Yang}}, \bibinfo {author} {\bibfnamefont {Z.}~\bibnamefont {Jia}}, \bibinfo
  {author} {\bibfnamefont {Y.}~\bibnamefont {Wu}}, \bibinfo {author}
  {\bibfnamefont {R.-C.}\ \bibnamefont {Xiao}}, \bibinfo {author}
  {\bibfnamefont {Z.~H.}\ \bibnamefont {Hang}}, \bibinfo {author}
  {\bibfnamefont {H.}~\bibnamefont {Jiang}}, \ and\ \bibinfo {author}
  {\bibfnamefont {X.}~\bibnamefont {Xie}},\ }\href@noop {} {\bibfield
  {journal} {\bibinfo  {journal} {Science Bulletin}\ }\textbf {\bibinfo
  {volume} {65}},\ \bibinfo {pages} {531 } (\bibinfo {year}
  {2020})}\BibitemShut {NoStop}%
\bibitem [{\citenamefont {Schindler}\ \emph
  {et~al.}(2018{\natexlab{b}})\citenamefont {Schindler}, \citenamefont {Wang},
  \citenamefont {Vergniory}, \citenamefont {Cook}, \citenamefont {Murani},
  \citenamefont {Sengupta}, \citenamefont {Kasumov}, \citenamefont {Deblock},
  \citenamefont {Jeon}, \citenamefont {Drozdov}, \citenamefont {Bouchiat},
  \citenamefont {Gu{\'e}ron}, \citenamefont {Yazdani}, \citenamefont
  {Bernevig},\ and\ \citenamefont {Neupert}}]{Schindler2018}%
  \BibitemOpen
  \bibfield  {author} {\bibinfo {author} {\bibfnamefont {F.}~\bibnamefont
  {Schindler}}, \bibinfo {author} {\bibfnamefont {Z.}~\bibnamefont {Wang}},
  \bibinfo {author} {\bibfnamefont {M.~G.}\ \bibnamefont {Vergniory}}, \bibinfo
  {author} {\bibfnamefont {A.~M.}\ \bibnamefont {Cook}}, \bibinfo {author}
  {\bibfnamefont {A.}~\bibnamefont {Murani}}, \bibinfo {author} {\bibfnamefont
  {S.}~\bibnamefont {Sengupta}}, \bibinfo {author} {\bibfnamefont {A.~Y.}\
  \bibnamefont {Kasumov}}, \bibinfo {author} {\bibfnamefont {R.}~\bibnamefont
  {Deblock}}, \bibinfo {author} {\bibfnamefont {S.}~\bibnamefont {Jeon}},
  \bibinfo {author} {\bibfnamefont {I.}~\bibnamefont {Drozdov}}, \bibinfo
  {author} {\bibfnamefont {H.}~\bibnamefont {Bouchiat}}, \bibinfo {author}
  {\bibfnamefont {S.}~\bibnamefont {Gu{\'e}ron}}, \bibinfo {author}
  {\bibfnamefont {A.}~\bibnamefont {Yazdani}}, \bibinfo {author} {\bibfnamefont
  {B.~A.}\ \bibnamefont {Bernevig}}, \ and\ \bibinfo {author} {\bibfnamefont
  {T.}~\bibnamefont {Neupert}},\ }\href@noop {} {\bibfield  {journal} {\bibinfo
   {journal} {Nat. Phys.}\ }\textbf {\bibinfo {volume} {14}},\ \bibinfo {pages}
  {918} (\bibinfo {year} {2018}{\natexlab{b}})}\BibitemShut {NoStop}%
\bibitem [{\citenamefont {Wang}\ \emph {et~al.}(2019)\citenamefont {Wang},
  \citenamefont {Wieder}, \citenamefont {Li}, \citenamefont {Yan},\ and\
  \citenamefont {Bernevig}}]{ZJWang_PRL2019}%
  \BibitemOpen
  \bibfield  {author} {\bibinfo {author} {\bibfnamefont {Z.}~\bibnamefont
  {Wang}}, \bibinfo {author} {\bibfnamefont {B.~J.}\ \bibnamefont {Wieder}},
  \bibinfo {author} {\bibfnamefont {J.}~\bibnamefont {Li}}, \bibinfo {author}
  {\bibfnamefont {B.}~\bibnamefont {Yan}}, \ and\ \bibinfo {author}
  {\bibfnamefont {B.~A.}\ \bibnamefont {Bernevig}},\ }\href@noop {} {\bibfield
  {journal} {\bibinfo  {journal} {Phys. Rev. Lett.}\ }\textbf {\bibinfo
  {volume} {123}},\ \bibinfo {pages} {186401} (\bibinfo {year}
  {2019})}\BibitemShut {NoStop}%
\bibitem [{\citenamefont {Yue}\ \emph {et~al.}(2019)\citenamefont {Yue},
  \citenamefont {Xu}, \citenamefont {Song}, \citenamefont {Weng}, \citenamefont
  {Lu}, \citenamefont {Fang},\ and\ \citenamefont {Dai}}]{Yue2019ws}%
  \BibitemOpen
  \bibfield  {author} {\bibinfo {author} {\bibfnamefont {C.}~\bibnamefont
  {Yue}}, \bibinfo {author} {\bibfnamefont {Y.}~\bibnamefont {Xu}}, \bibinfo
  {author} {\bibfnamefont {Z.}~\bibnamefont {Song}}, \bibinfo {author}
  {\bibfnamefont {H.}~\bibnamefont {Weng}}, \bibinfo {author} {\bibfnamefont
  {Y.-M.}\ \bibnamefont {Lu}}, \bibinfo {author} {\bibfnamefont
  {C.}~\bibnamefont {Fang}}, \ and\ \bibinfo {author} {\bibfnamefont
  {X.}~\bibnamefont {Dai}},\ }\href@noop {} {\bibfield  {journal} {\bibinfo
  {journal} {Nat. Phys.}\ }\textbf {\bibinfo {volume} {15}},\ \bibinfo {pages}
  {577} (\bibinfo {year} {2019})}\BibitemShut {NoStop}%
\bibitem [{\citenamefont {Xu}\ \emph {et~al.}(2019)\citenamefont {Xu},
  \citenamefont {Song}, \citenamefont {Wang}, \citenamefont {Weng},\ and\
  \citenamefont {Dai}}]{XuYF2019}%
  \BibitemOpen
  \bibfield  {author} {\bibinfo {author} {\bibfnamefont {Y.}~\bibnamefont
  {Xu}}, \bibinfo {author} {\bibfnamefont {Z.}~\bibnamefont {Song}}, \bibinfo
  {author} {\bibfnamefont {Z.}~\bibnamefont {Wang}}, \bibinfo {author}
  {\bibfnamefont {H.}~\bibnamefont {Weng}}, \ and\ \bibinfo {author}
  {\bibfnamefont {X.}~\bibnamefont {Dai}},\ }\href@noop {} {\bibfield
  {journal} {\bibinfo  {journal} {Phys. Rev. Lett.}\ }\textbf {\bibinfo
  {volume} {122}},\ \bibinfo {pages} {256402} (\bibinfo {year}
  {2019})}\BibitemShut {NoStop}%
\bibitem [{\citenamefont {Zhang}\ \emph {et~al.}(2020)\citenamefont {Zhang},
  \citenamefont {Wu},\ and\ \citenamefont {Das~Sarma}}]{MnBiTe_PRL}%
  \BibitemOpen
  \bibfield  {author} {\bibinfo {author} {\bibfnamefont {R.-X.}\ \bibnamefont
  {Zhang}}, \bibinfo {author} {\bibfnamefont {F.}~\bibnamefont {Wu}}, \ and\
  \bibinfo {author} {\bibfnamefont {S.}~\bibnamefont {Das~Sarma}},\ }\href@noop
  {} {\bibfield  {journal} {\bibinfo  {journal} {Phys. Rev. Lett.}\ }\textbf
  {\bibinfo {volume} {124}},\ \bibinfo {pages} {136407} (\bibinfo {year}
  {2020})}\BibitemShut {NoStop}%
\bibitem [{\citenamefont {Sheng}\ \emph {et~al.}(2019)\citenamefont {Sheng},
  \citenamefont {Chen}, \citenamefont {Liu}, \citenamefont {Chen},
  \citenamefont {Yu}, \citenamefont {Zhao},\ and\ \citenamefont
  {Yang}}]{GDY_PRL2019}%
  \BibitemOpen
  \bibfield  {author} {\bibinfo {author} {\bibfnamefont {X.-L.}\ \bibnamefont
  {Sheng}}, \bibinfo {author} {\bibfnamefont {C.}~\bibnamefont {Chen}},
  \bibinfo {author} {\bibfnamefont {H.}~\bibnamefont {Liu}}, \bibinfo {author}
  {\bibfnamefont {Z.}~\bibnamefont {Chen}}, \bibinfo {author} {\bibfnamefont
  {Z.-M.}\ \bibnamefont {Yu}}, \bibinfo {author} {\bibfnamefont {Y.~X.}\
  \bibnamefont {Zhao}}, \ and\ \bibinfo {author} {\bibfnamefont {S.~A.}\
  \bibnamefont {Yang}},\ }\href@noop {} {\bibfield  {journal} {\bibinfo
  {journal} {Phys. Rev. Lett.}\ }\textbf {\bibinfo {volume} {123}},\ \bibinfo
  {pages} {256402} (\bibinfo {year} {2019})}\BibitemShut {NoStop}%
\bibitem [{\citenamefont {Lee}\ \emph {et~al.}(2020)\citenamefont {Lee},
  \citenamefont {Kim}, \citenamefont {Ahn},\ and\ \citenamefont
  {Yang}}]{Lee_GDY}%
  \BibitemOpen
  \bibfield  {author} {\bibinfo {author} {\bibfnamefont {E.}~\bibnamefont
  {Lee}}, \bibinfo {author} {\bibfnamefont {R.}~\bibnamefont {Kim}}, \bibinfo
  {author} {\bibfnamefont {J.}~\bibnamefont {Ahn}}, \ and\ \bibinfo {author}
  {\bibfnamefont {B.-J.}\ \bibnamefont {Yang}},\ }\href@noop {} {\bibfield
  {journal} {\bibinfo  {journal} {npj Quantum Mater.}\ }\textbf {\bibinfo
  {volume} {5}},\ \bibinfo {pages} {1} (\bibinfo {year} {2020})}\BibitemShut
  {NoStop}%
\bibitem [{\citenamefont {Zhao}\ and\ \citenamefont
  {Lu}(2017)}]{RealCN_YXZhao}%
  \BibitemOpen
  \bibfield  {author} {\bibinfo {author} {\bibfnamefont {Y.~X.}\ \bibnamefont
  {Zhao}}\ and\ \bibinfo {author} {\bibfnamefont {Y.}~\bibnamefont {Lu}},\
  }\href@noop {} {\bibfield  {journal} {\bibinfo  {journal} {Phys. Rev. Lett.}\
  }\textbf {\bibinfo {volume} {118}},\ \bibinfo {pages} {056401} (\bibinfo
  {year} {2017})}\BibitemShut {NoStop}%
\bibitem [{\citenamefont {Kang}\ \emph {et~al.}(2011)\citenamefont {Kang},
  \citenamefont {Li}, \citenamefont {Wu}, \citenamefont {Li},\ and\
  \citenamefont {Xia}}]{JunKangJPCC}%
  \BibitemOpen
  \bibfield  {author} {\bibinfo {author} {\bibfnamefont {J.}~\bibnamefont
  {Kang}}, \bibinfo {author} {\bibfnamefont {J.}~\bibnamefont {Li}}, \bibinfo
  {author} {\bibfnamefont {F.}~\bibnamefont {Wu}}, \bibinfo {author}
  {\bibfnamefont {S.-S.}\ \bibnamefont {Li}}, \ and\ \bibinfo {author}
  {\bibfnamefont {J.-B.}\ \bibnamefont {Xia}},\ }\href@noop {} {\bibfield
  {journal} {\bibinfo  {journal} {J. Phys. Chem. C}\ ,\ \bibinfo {pages}
  {20466–20470}} (\bibinfo {year} {2011})}\BibitemShut {NoStop}%
\bibitem [{\citenamefont {Zhao}\ \emph {et~al.}(2016)\citenamefont {Zhao},
  \citenamefont {Schnyder},\ and\ \citenamefont
  {Wang}}]{ZhaoYX_Schhyder_PRL2016}%
  \BibitemOpen
  \bibfield  {author} {\bibinfo {author} {\bibfnamefont {Y.~X.}\ \bibnamefont
  {Zhao}}, \bibinfo {author} {\bibfnamefont {A.~P.}\ \bibnamefont {Schnyder}},
  \ and\ \bibinfo {author} {\bibfnamefont {Z.~D.}\ \bibnamefont {Wang}},\
  }\href@noop {} {\bibfield  {journal} {\bibinfo  {journal} {Phys. Rev. Lett.}\
  }\textbf {\bibinfo {volume} {116}},\ \bibinfo {pages} {156402} (\bibinfo
  {year} {2016})}\BibitemShut {NoStop}%
\bibitem [{\citenamefont {Ahn}\ \emph {et~al.}(2019)\citenamefont {Ahn},
  \citenamefont {Park}, \citenamefont {Kim}, \citenamefont {Kim},\ and\
  \citenamefont {Yang}}]{BJYang_CPB}%
  \BibitemOpen
  \bibfield  {author} {\bibinfo {author} {\bibfnamefont {J.}~\bibnamefont
  {Ahn}}, \bibinfo {author} {\bibfnamefont {S.}~\bibnamefont {Park}}, \bibinfo
  {author} {\bibfnamefont {D.}~\bibnamefont {Kim}}, \bibinfo {author}
  {\bibfnamefont {Y.}~\bibnamefont {Kim}}, \ and\ \bibinfo {author}
  {\bibfnamefont {B.-J.}\ \bibnamefont {Yang}},\ }\href@noop {} {\bibfield
  {journal} {\bibinfo  {journal} {Chin. Phys. B}\ }\textbf {\bibinfo {volume}
  {28}},\ \bibinfo {eid} {117101} (\bibinfo {year} {2019})}\BibitemShut
  {NoStop}%
\bibitem [{\citenamefont {Zak}(1989)}]{Zak_PRL1989}%
  \BibitemOpen
  \bibfield  {author} {\bibinfo {author} {\bibfnamefont {J.}~\bibnamefont
  {Zak}},\ }\href@noop {} {\bibfield  {journal} {\bibinfo  {journal} {Phys.
  Rev. Lett.}\ }\textbf {\bibinfo {volume} {62}},\ \bibinfo {pages} {2747}
  (\bibinfo {year} {1989})}\BibitemShut {NoStop}%
\bibitem [{\citenamefont {Milnor}\ and\ \citenamefont
  {Stasheff}(2016)}]{milnor2016characteristic}%
  \BibitemOpen
  \bibfield  {author} {\bibinfo {author} {\bibfnamefont {J.}~\bibnamefont
  {Milnor}}\ and\ \bibinfo {author} {\bibfnamefont {J.~D.}\ \bibnamefont
  {Stasheff}},\ }\href@noop {} {\emph {\bibinfo {title} {Characteristic
  Classes}}},\ Vol.~\bibinfo {volume} {76}\ (\bibinfo  {publisher} {Princeton
  university press},\ \bibinfo {year} {2016})\BibitemShut {NoStop}%
\bibitem [{\citenamefont {Ahn}\ \emph {et~al.}(2018)\citenamefont {Ahn},
  \citenamefont {Kim}, \citenamefont {Kim},\ and\ \citenamefont
  {Yang}}]{JBYang_PRL2018}%
  \BibitemOpen
  \bibfield  {author} {\bibinfo {author} {\bibfnamefont {J.}~\bibnamefont
  {Ahn}}, \bibinfo {author} {\bibfnamefont {D.}~\bibnamefont {Kim}}, \bibinfo
  {author} {\bibfnamefont {Y.}~\bibnamefont {Kim}}, \ and\ \bibinfo {author}
  {\bibfnamefont {B.-J.}\ \bibnamefont {Yang}},\ }\href@noop {} {\bibfield
  {journal} {\bibinfo  {journal} {Phys. Rev. Lett.}\ }\textbf {\bibinfo
  {volume} {121}},\ \bibinfo {pages} {106403} (\bibinfo {year}
  {2018})}\BibitemShut {NoStop}%
\bibitem [{\citenamefont {Yu}\ \emph {et~al.}(2011)\citenamefont {Yu},
  \citenamefont {Qi}, \citenamefont {Bernevig}, \citenamefont {Fang},\ and\
  \citenamefont {Dai}}]{YuRui_PRB2011}%
  \BibitemOpen
  \bibfield  {author} {\bibinfo {author} {\bibfnamefont {R.}~\bibnamefont
  {Yu}}, \bibinfo {author} {\bibfnamefont {X.~L.}\ \bibnamefont {Qi}}, \bibinfo
  {author} {\bibfnamefont {A.}~\bibnamefont {Bernevig}}, \bibinfo {author}
  {\bibfnamefont {Z.}~\bibnamefont {Fang}}, \ and\ \bibinfo {author}
  {\bibfnamefont {X.}~\bibnamefont {Dai}},\ }\href@noop {} {\bibfield
  {journal} {\bibinfo  {journal} {Phys. Rev. B}\ }\textbf {\bibinfo {volume}
  {84}},\ \bibinfo {pages} {075119} (\bibinfo {year} {2011})}\BibitemShut
  {NoStop}%
\bibitem [{\citenamefont {Shen}(2012)}]{ShunQingShen_TI}%
  \BibitemOpen
  \bibfield  {author} {\bibinfo {author} {\bibfnamefont {S.-Q.}\ \bibnamefont
  {Shen}},\ }\href@noop {} {\emph {\bibinfo {title} {Topological Insulators}}}\
  (\bibinfo  {publisher} {Springer, Berlin},\ \bibinfo {year}
  {2012})\BibitemShut {NoStop}%
\bibitem [{\citenamefont {Jackiw}\ and\ \citenamefont
  {Rebbi}(1976)}]{Jackiw_PRD}%
  \BibitemOpen
  \bibfield  {author} {\bibinfo {author} {\bibfnamefont {R.}~\bibnamefont
  {Jackiw}}\ and\ \bibinfo {author} {\bibfnamefont {C.}~\bibnamefont {Rebbi}},\
  }\href@noop {} {\bibfield  {journal} {\bibinfo  {journal} {Phys. Rev. D}\
  }\textbf {\bibinfo {volume} {13}},\ \bibinfo {pages} {3398} (\bibinfo {year}
  {1976})}\BibitemShut {NoStop}%
\bibitem [{\citenamefont {Kresse}\ and\ \citenamefont
  {Hafner}(1994)}]{Kresse1994}%
  \BibitemOpen
  \bibfield  {author} {\bibinfo {author} {\bibfnamefont {G.}~\bibnamefont
  {Kresse}}\ and\ \bibinfo {author} {\bibfnamefont {J.}~\bibnamefont
  {Hafner}},\ }\href@noop {} {\bibfield  {journal} {\bibinfo  {journal} {Phys.
  Rev. B}\ }\textbf {\bibinfo {volume} {49}},\ \bibinfo {pages} {14251}
  (\bibinfo {year} {1994})}\BibitemShut {NoStop}%
\bibitem [{\citenamefont {Kresse}\ and\ \citenamefont
  {Furthm\"uller}(1996)}]{Kresse1996}%
  \BibitemOpen
  \bibfield  {author} {\bibinfo {author} {\bibfnamefont {G.}~\bibnamefont
  {Kresse}}\ and\ \bibinfo {author} {\bibfnamefont {J.}~\bibnamefont
  {Furthm\"uller}},\ }\href@noop {} {\bibfield  {journal} {\bibinfo  {journal}
  {Phys. Rev. B}\ }\textbf {\bibinfo {volume} {54}},\ \bibinfo {pages} {11169}
  (\bibinfo {year} {1996})}\BibitemShut {NoStop}%
\bibitem [{\citenamefont {Bl\"ochl}(1994)}]{PAW}%
  \BibitemOpen
  \bibfield  {author} {\bibinfo {author} {\bibfnamefont {P.~E.}\ \bibnamefont
  {Bl\"ochl}},\ }\href@noop {} {\bibfield  {journal} {\bibinfo  {journal}
  {Phys. Rev. B}\ }\textbf {\bibinfo {volume} {50}},\ \bibinfo {pages} {17953}
  (\bibinfo {year} {1994})}\BibitemShut {NoStop}%
\bibitem [{\citenamefont {Perdew}\ \emph {et~al.}(1996)\citenamefont {Perdew},
  \citenamefont {Burke},\ and\ \citenamefont {Ernzerhof}}]{PBE}%
  \BibitemOpen
  \bibfield  {author} {\bibinfo {author} {\bibfnamefont {J.~P.}\ \bibnamefont
  {Perdew}}, \bibinfo {author} {\bibfnamefont {K.}~\bibnamefont {Burke}}, \
  and\ \bibinfo {author} {\bibfnamefont {M.}~\bibnamefont {Ernzerhof}},\
  }\href@noop {} {\bibfield  {journal} {\bibinfo  {journal} {Phys. Rev. Lett.}\
  }\textbf {\bibinfo {volume} {77}},\ \bibinfo {pages} {3865} (\bibinfo {year}
  {1996})}\BibitemShut {NoStop}%
\bibitem [{\citenamefont {Monkhorst}\ and\ \citenamefont
  {Pack}(1976)}]{PhysRevB.13.5188}%
  \BibitemOpen
  \bibfield  {author} {\bibinfo {author} {\bibfnamefont {H.~J.}\ \bibnamefont
  {Monkhorst}}\ and\ \bibinfo {author} {\bibfnamefont {J.~D.}\ \bibnamefont
  {Pack}},\ }\href@noop {} {\bibfield  {journal} {\bibinfo  {journal} {Phys.
  Rev. B}\ }\textbf {\bibinfo {volume} {13}},\ \bibinfo {pages} {5188}
  (\bibinfo {year} {1976})}\BibitemShut {NoStop}%
\bibitem [{\citenamefont {Becke}\ and\ \citenamefont {Johnson}(2006)}]{mBJ}%
  \BibitemOpen
  \bibfield  {author} {\bibinfo {author} {\bibfnamefont {A.~D.}\ \bibnamefont
  {Becke}}\ and\ \bibinfo {author} {\bibfnamefont {E.~R.}\ \bibnamefont
  {Johnson}},\ }\href@noop {} {\bibfield  {journal} {\bibinfo  {journal} {J.
  Chem. Phys.}\ }\textbf {\bibinfo {volume} {124}},\ \bibinfo {pages} {221101}
  (\bibinfo {year} {2006})}\BibitemShut {NoStop}%
\bibitem [{\citenamefont {Marzari}\ and\ \citenamefont
  {Vanderbilt}(1997)}]{Marzari1997}%
  \BibitemOpen
  \bibfield  {author} {\bibinfo {author} {\bibfnamefont {N.}~\bibnamefont
  {Marzari}}\ and\ \bibinfo {author} {\bibfnamefont {D.}~\bibnamefont
  {Vanderbilt}},\ }\href@noop {} {\bibfield  {journal} {\bibinfo  {journal}
  {Phys. Rev. B}\ }\textbf {\bibinfo {volume} {56}},\ \bibinfo {pages} {12847}
  (\bibinfo {year} {1997})}\BibitemShut {NoStop}%
\bibitem [{\citenamefont {Souza}\ \emph {et~al.}(2001)\citenamefont {Souza},
  \citenamefont {Marzari},\ and\ \citenamefont {Vanderbilt}}]{Souza2001}%
  \BibitemOpen
  \bibfield  {author} {\bibinfo {author} {\bibfnamefont {I.}~\bibnamefont
  {Souza}}, \bibinfo {author} {\bibfnamefont {N.}~\bibnamefont {Marzari}}, \
  and\ \bibinfo {author} {\bibfnamefont {D.}~\bibnamefont {Vanderbilt}},\
  }\href@noop {} {\bibfield  {journal} {\bibinfo  {journal} {Phys. Rev. B}\
  }\textbf {\bibinfo {volume} {65}},\ \bibinfo {pages} {035109} (\bibinfo
  {year} {2001})}\BibitemShut {NoStop}%
\bibitem [{\citenamefont {Wu}\ \emph {et~al.}(2018)\citenamefont {Wu},
  \citenamefont {Zhang}, \citenamefont {Song}, \citenamefont {Troyer},\ and\
  \citenamefont {Soluyanov}}]{Wu2017}%
  \BibitemOpen
  \bibfield  {author} {\bibinfo {author} {\bibfnamefont {Q.}~\bibnamefont
  {Wu}}, \bibinfo {author} {\bibfnamefont {S.}~\bibnamefont {Zhang}}, \bibinfo
  {author} {\bibfnamefont {H.-F.}\ \bibnamefont {Song}}, \bibinfo {author}
  {\bibfnamefont {M.}~\bibnamefont {Troyer}}, \ and\ \bibinfo {author}
  {\bibfnamefont {A.~A.}\ \bibnamefont {Soluyanov}},\ }\href@noop {} {\bibfield
   {journal} {\bibinfo  {journal} {Comput. Phys. Commun.}\ }\textbf {\bibinfo
  {volume} {224}},\ \bibinfo {pages} {405 } (\bibinfo {year}
  {2018})}\BibitemShut {NoStop}%
\bibitem [{\citenamefont {Sancho}\ \emph {et~al.}(1984)\citenamefont {Sancho},
  \citenamefont {Sancho},\ and\ \citenamefont {Rubio}}]{Green1984}%
  \BibitemOpen
  \bibfield  {author} {\bibinfo {author} {\bibfnamefont {M.~P.~L.}\
  \bibnamefont {Sancho}}, \bibinfo {author} {\bibfnamefont {J.~M.~L.}\
  \bibnamefont {Sancho}}, \ and\ \bibinfo {author} {\bibfnamefont
  {J.}~\bibnamefont {Rubio}},\ }\href@noop {} {\bibfield  {journal} {\bibinfo
  {journal} {J. Phys. F}\ }\textbf {\bibinfo {volume} {14}},\ \bibinfo {pages}
  {1205} (\bibinfo {year} {1984})}\BibitemShut {NoStop}%
\bibitem [{\citenamefont {Sancho}\ \emph {et~al.}(1985)\citenamefont {Sancho},
  \citenamefont {Sancho}, \citenamefont {Sancho},\ and\ \citenamefont
  {Rubio}}]{Green1985}%
  \BibitemOpen
  \bibfield  {author} {\bibinfo {author} {\bibfnamefont {M.~P.~L.}\
  \bibnamefont {Sancho}}, \bibinfo {author} {\bibfnamefont {J.~M.~L.}\
  \bibnamefont {Sancho}}, \bibinfo {author} {\bibfnamefont {J.~M.~L.}\
  \bibnamefont {Sancho}}, \ and\ \bibinfo {author} {\bibfnamefont
  {J.}~\bibnamefont {Rubio}},\ }\href@noop {} {\bibfield  {journal} {\bibinfo
  {journal} {J. Phys. F}\ }\textbf {\bibinfo {volume} {15}},\ \bibinfo {pages}
  {851} (\bibinfo {year} {1985})}\BibitemShut {NoStop}%
\end{thebibliography}%


\end{document}